\documentclass[prl,twocolumn,amsmath,amssymb,showpacs,aps]{revtex4-2}
\usepackage[utf8]{inputenc}
\usepackage{color}
\usepackage{latexsym}

\usepackage{float}
\usepackage{dsfont}
\usepackage[dvipsnames]{xcolor}
\usepackage[english]{babel}
\usepackage{cancel}

\usepackage{bm, mathtools}
\usepackage{graphicx}
\usepackage{dcolumn}
\usepackage{times}
\usepackage[english]{babel}
\usepackage[T1]{fontenc}

\usepackage{cellspace}%
\usepackage{makecell}
\setcellgapes{3pt}

\newcommand{\dotnice}[1]{\overset{\,\bm.}{#1}{\vphantom{#1}}}
\newcommand{\ddotnice}[1]{\overset{\,\bm.\bm.}{#1}{\vphantom{#1}}}

\renewcommand{\vec}{\bm}
\newcommand{\smean}[1]{\langle{#1}\rangle}
\newcommand{\mean}[1]{\left\langle {#1} \right\rangle}

\newcommand{\dd}{\mathrm{d}}
\renewcommand{\AA}{A}

\newcommand{\Cov}[2]{{{\mathrm{Cov}}{\left({#1}\,, {#2} \right)}}}
\newcommand{\sym}{{^{i	\rightleftarrows j}}}
\newcommand{\Sym}[1]{{{\left[{#1}\right]}\sym}}

\definecolor{linkcolor}{rgb}{0,0,0.6} 

\newcommand{\inflow}{\mathcal{G}}
\newcommand{\traffic}{\mathcal{T}}

\newcommand{\tcr}[1]{\textcolor{black}{#1}}

\usepackage[pdftex,colorlinks=true, pdfstartview=FitV, linkcolor= linkcolor, citecolor= linkcolor, urlcolor= linkcolor, hyperindex=true,hyperfigures=true]{hyperref} 

\begin{document}

\title{Force-Free Kinetic Inference of Entropy Production}

\author{I. Di Terlizzi}

\affiliation{Max Planck Institute for the Physics of Complex Systems, N{\"o}thnitzer Stra{\ss}e 38, 01187, Dresden, Germany}


\begin{abstract}
\noindent 
Estimating entropy production, which quantifies irreversibility and energy dissipation, remains a significant challenge despite its central role in nonequilibrium physics. We propose a novel method for estimating the mean entropy production rate $\sigma$ that relies solely on position traces, bypassing the need for flux or microscopic force measurements. Starting from a recently introduced variance sum rule, we express $\sigma$ in terms of measurable steady-state correlation functions which we link to previously studied kinetic quantities, known as traffic and inflow rate. Under realistic constraints of limited access to dynamical degrees of freedom, we derive efficient bounds on $\sigma$ by leveraging the information contained in the system's traffic, enabling partial but meaningful estimates of $\sigma$. We benchmark our results across several orders of magnitude in $\sigma$ using two models: a linear stochastic system and a nonlinear model for spontaneous hair-bundle oscillations. Our approach offers a practical and versatile framework for investigating entropy production in nonequilibrium systems.

\end{abstract}

\maketitle

Stochastic modeling is essential for describing complex phenomena characterized by randomness and fluctuations. Their breaking of time-reversal symmetry and energy dissipation are quantified by the \textit{entropy production}~\cite{maes03_1,seifert2012stochastic,peliti2021stochastic,GalCohenFluc,fluc1,kurchan1998fluctuation}.  
The irreversibility of stochastic traces
can be directly used to infer the entropy production rate $\sigma$ in a nonequilibrium steady state (NESS)~\cite{andrieux2007entropy,li2019quantifying,martinez2019inferring,ro2022model}. However, estimating $\sigma$ exploiting violations of time-reversal symmetry typically requires observing stochastic trajectories whose length scales exponentially with $\sigma$, posing practical limitations. Recent advances offer an alternative approach by bounding entropy production using information-theoretic methods, leading to the development of thermodynamic uncertainty relations (TURs)~\cite{bar15,pietzonka2017finite,macieszczak2018unified,busiello2019hyperaccurate,hasegawa2019fluctuation,hasegawa2019uncertainty,horowitz2020thermodynamic,otsubo2020estimating,falasco2020unifying,Di_Terlizzi_Mem_TUR,manikandan2020inferring,van2020entropy,shiraishi2021optimal,van2022unified,manikandan2021quantitative,otsubo2022estimating,dieball2023direct,pietzonka2024thermodynamic,lucente2025conceptual}. Alternative methods require perturbation experiments to exploit the violation of the fluctuation-dissipation theorem ~\cite{turlier2016equilibrium,wang2018inferring} by directly applying the Harada-Sasa relation \cite{har05}. Another possibility could be to directly measure microscopic forces or probability fluxes and use their relation with dissipation~\cite{sekimoto1998langevin,Gnesotto_2018}, but this is challenging in most cases. Moreover, in many realistic scenarios, the system under study can only be partially observed, which further complicates the inference process. For discrete systems, this typically corresponds to observing coarse-grained or lumped mesostates, which aggregate numerous microstates whose individual transitions remain unobservable \cite{roldan2012entropy,bisker2017hierarchical,busiello2019entropy,skinner2021estimating,skinner2021improved,teza2020exact,ghosal2023entropy,harunari2022learn,van2022thermodynamic,blom2024milestoning,baiesi2024effective,fritz2025entropy}. For continuous processes, limited information may result from spatial coarse-graining \cite{gingrich2017inferring,dieball2022mathematical,dieball2022coarse,ghosal2022inferring} or from the partial observation of a subset of the Markovian degrees of freedom \cite{roldan2010estimating,gnesotto2020learning,roldan2021quantifying,busiello2024unraveling}. While central to understanding irreversible processes, estimating entropy production remains a challenging and active area of ongoing research.

In this Letter, we present a framework to estimate $\sigma$ directly from stochastic trajectories, without requiring force measurements or inference of irreversibility, and applicable even under partial observation. We consider steady-state dynamics of multiple degrees of freedom (DOFs) $\{x^i_t\}$ governed by the overdamped Langevin equation,
\begin{equation} \label{LE}
\dot{\pmb{x}}_t = \pmb{\mu}\pmb{F}(\pmb{x}_t) + \sqrt{2\pmb{D}}\,\pmb{\xi}_t \, ,
\end{equation}
with potentially non-conservative forces $\pmb{F}(\pmb{x}_t)$ and white noise with mean $\langle \xi^i_{t} \rangle=0$ and covariance $\langle \xi^i_{t}\xi^j_{s} \rangle=\delta^{ij}\delta(t-s)$. {This effective description is valid in the overdamped regime, which is typical for mesoscopic systems where inertia is negligible compared to viscous damping, as commonly encountered in biological and soft matter systems. More generally, such a continuous description may emerge as a coarse-grained limit of an underlying discrete stochastic process with fast microscopic dynamics. Under broad conditions, these processes admit a diffusion approximation, ensuring that the inferred entropy production still provides a meaningful lower bound on the total dissipation \cite{busiello2019entropy}.} The system may be in contact with different thermal baths $T^i$ encoded in a diagonal temperature matrix $\pmb{T}$. The mobility matrix $\pmb{\mu}$ is related to the diffusion matrix by the Einstein relation $\pmb{D}=k_{\rm B}\pmb{T}\pmb{\mu}$. According to stochastic energetics
\cite{sekimoto1998langevin}, in a NESS $\sigma$ corresponds to the amount of heat dissipated in the environment per unit time,
\begin{equation}
\label{sekimoto}
\sigma = \sum_i\frac{\langle \Delta Q^i \rangle }{k_{\rm B} T^i} = \sum_i\frac{1}{k_{\rm B} T^i} \langle F^i_t\circ \dot{x}^i_t\rangle \;,
\end{equation}
where $\langle \Delta Q^i \rangle$ is the average heat injected into the $i^{\rm th}$ heat bath at temperature $T^i$ and $\circ$ denotes the Stratonovich product. This formula often exposes two main empirical problems: (i) it requires measuring the forces acting on the relevant DOFs (which is challenging in most cases), and (ii) it often clashes with the impossibility of observing all relevant DOFs. 

In \cite{VSR,VSR_solved_models} we derived a formula for $\sigma$ starting from the variance sum rule (VSR) involving the second derivative of the position correlation function and the covariances of microscopic forces. Expressed in terms of correlation functions and following Einstein's summation convention, it takes the form:
\begin{equation}\label{sigma_VSR_old}
    \sigma = \frac{1}{2} (D^{\rm -1})^{ij}\left(- 
 \ddotnice{C}_{x}^{\, ij}(0)
 + \,C^{\, ij}_{\mu F}(0)\right)
\end{equation}
where $\pmb{C}^{\,ij}_O(t)\equiv \langle O^i_t O^j_0 \rangle-\langle O^i_t \rangle\langle O^i_0 \rangle$ is the connected correlation matrix for the observable $\pmb{O}$ and $\ddotnice{\pmb{C}}_O(0)\equiv \partial_t^2\,\pmb{C}_O(t)|_{t=0^+}$ is its second derivative evaluated at zero. Eq. \eqref{sigma_VSR_old} shows that the curvatures of the position correlation functions at short times are highly informative about dissipation. Still, the difficulties in estimating mobility and forces in experiments hinder the direct application of this formula. For example, in \cite{VSR}, we analyzed experimental measurements of red blood cell flickering, where only a single degree of freedom was observable and direct measurements of cellular forces were not accessible. To overcome these limitations, a reduced form of the VSR was employed to fit the experimental data, and, similar to \cite{roldan2021quantifying,Tucci22}, modeling was necessary to compensate for the lack of information about the underlying system. 

Starting from the VSR, we derive a new formula for $\sigma$ (see~\eqref{main_eq} below) that does not rely on the evaluation of forces. First, we deal with cases where all relevant DOFs are visible, even if forces are not directly measurable. When some DOFs are not detectable, we derive bounds on $\sigma$, enabling a partial yet informative estimation process. {Rather than relying on microscopic forces, our method estimates effective forces from the experimentally accessible probability density function (PDF) $p(\pmb{x}_t)$. In a NESS, the effective potential is $\phi(\pmb{x}_t) = -\log p(\pmb{x}_t) + C$, with $C$ a constant dependent on the PDF's normalization, and its gradient, scaled by $-k_{\rm B} T$, acts as an effective force. We thus reformulate Eq.~\eqref{sigma_VSR_old} in terms of effective forces using the \textit{mean local velocity} $\pmb{\nu}(\pmb{x}_t)$, obtained from the Fokker-Planck equation associated with Eq.~\eqref{LE},}
\begin{align}\label{eq:Anu_main} \pmb{\nu}(\pmb{x}_t) = \pmb{\mu} \pmb{F}(\pmb{x}_t) + \pmb{D} \nabla \phi(\pmb{x}_t) \, , 
\end{align}
{which connects real forces $\pmb{F}(\pmb{x}_t)$ to $-\nabla \phi(\pmb{x}_t)$. This leads, as shown in Section S1 of~\cite{SuppMat}, to a novel expression for $\sigma$:}
\begin{equation}\label{main_eq}
    \sigma = - (D^{\rm -1})^{ij}\, 
 \ddotnice{C}_{x}^{\, ij}(0)
 + D^{ij}\,C^{\, ij}_{\nabla\phi}(0)\, ,
\end{equation}
{where all terms are directly inferred from position traces.} Indeed, the diffusion matrix, $D^{ij}$, can be directly extracted from the position correlation matrix, $C^{\, ij}_x(t)$, via 
\begin{equation}
    D^{ij} = -\dotnice{C}^{\,(ij)}_x(0) \, ,
\end{equation}
where $C^{(ij)}= (C^{\,ij}+C^{\,ji})/2$
denotes the symmetrized matrix and $\dotnice{\pmb{C}}_x(0)\equiv \partial_t\,\pmb{C}_x(t)|_{t=0^+}$ (see Section S5 in \cite{SuppMat}
). {The linear short-time behavior of $\pmb{C}_x(t)$ is a signature of overdamped dynamics, as underdamped systems exhibit a ballistic (quadratic) onset instead (see Section S5A); this provides a direct way to verify that the overdamped approximation is appropriate for the system under study.} The second term on the right-hand side of Eq.~\eqref{main_eq} corresponds to the inflow rate 
\begin{equation}
    \inflow = D^{ij}\,C^{\, ij}_{\nabla\phi}(0) \, ,
\end{equation}
a time-symmetric quantity obeying fluctuation relations \cite{fal15c} and recently used to study the information content of stochastic traces \cite{frishman2020learning}. $\inflow$ can be determined from effective force measurements, which involves estimating $p(\pmb{x}_t)$ using a kernel density estimator and computing the gradient $\nabla\phi(\pmb{x}_t)$ with standard numerical tools, see Section S11 in \cite{SuppMat} 
for more details. The connection between $\sigma$ and $\inflow$ was first established in \cite{maes2008steady}, with the traffic $\traffic$, a key kinetic quantity also commonly referred to as \textit{dynamical activity} or \textit{frenesy} \cite{baiesi2018life,dit19,MAES20201}, providing the link between the two. This relationship is given by
\begin{align}\label{main_eq_exp}
\sigma = 4\mathcal{T} + \inflow \, ,
\end{align}
and, together with \eqref{main_eq}, leads to the identification
\begin{align}\label{traffic_new}
\traffic = - (D^{\rm -1})^{ij}\,
\ddotnice{C}_{x}^{\, ij}(0)/4 \,.
\end{align}
The traffic represents the symmetric component of the stochastic action associated with the Langevin process. While its original definition \cite{maes2008steady} (see Eq. S17 in \cite{SuppMat}) is based on microscopic forces, the new formulation \eqref{traffic_new} enables direct estimation from position measurements. To summarize, since both $\inflow$ and $\traffic$ can be expressed in terms of derivatives of position correlation functions $\pmb{C}_x(t)$ and effective force covariances $\pmb{C}_{\nabla \phi}(0)$, both terms in \eqref{main_eq_exp} can be accurately evaluated from position traces. {In the following examples, we show that $\traffic$ typically tracks the behavior of $\sigma$ more closely than $\inflow$. The inflow rate $\inflow$ reflects how the steady-state distribution is shaped by the forces, acting as a measure of local compression or spreading of probability, and is more reflective of static features of the system. In contrast, $\traffic$ encodes time-symmetric dynamical fluctuations and reflects local kinetic activity. As shown in Section~S3 of~\cite{SuppMat}, this distinction becomes crucial in strongly driven regimes, where $\traffic$ continues to grow with dissipation while $\inflow$ remains comparatively small.}

To illustrate our approach, we consider two representative systems: (i) the analytically tractable Ornstein–Uhlenbeck process, and (ii) a nonlinear model of spontaneous hair-bundle oscillations in bullfrog ears \cite{hair0}. While the former enables full theoretical analysis, the latter highlights biological applicability. Across both, we identify parameter regimes where entropy production $\sigma$ varies significantly, and show that this variability is largely captured by the traffic $\traffic$, which remains accessible even under partial observation, unlike $\inflow$.

\begin{figure*}[ht!]
 	\centering
  \includegraphics[width=1\textwidth]{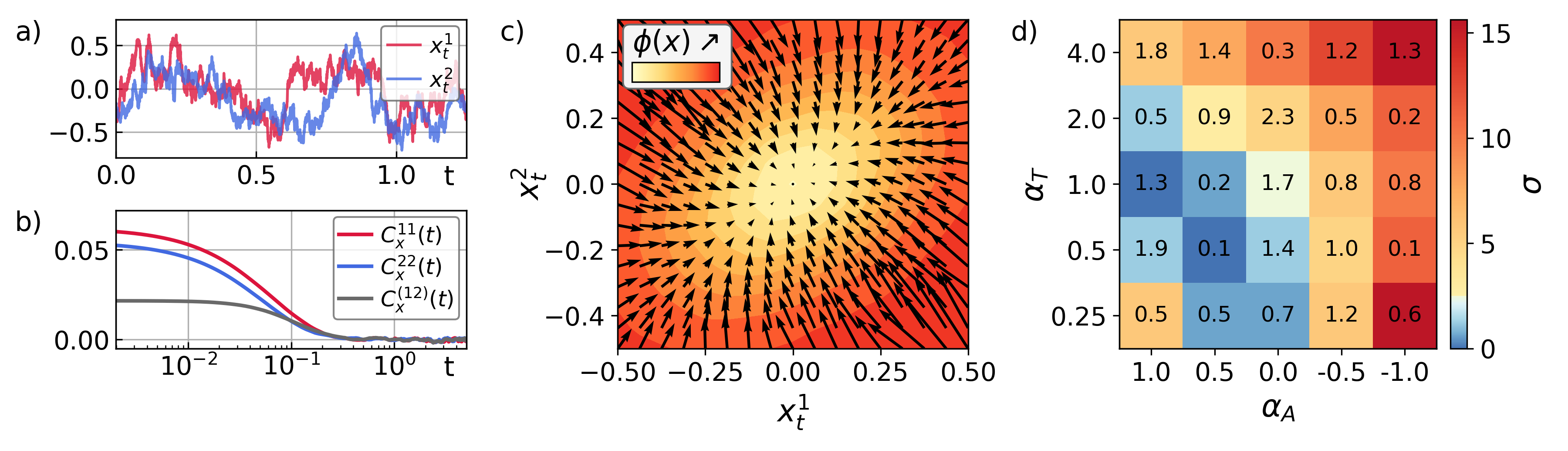}
	\caption{a) Stochastic traces of $x^1_t$ and $x^2_t$, sampled at $4\,{\rm kHz}$ over a duration of 500~s.
    b) Corresponding correlation functions for both degrees of freedom (DOFs). c) Effective forces $-\nabla \phi$ (black arrows) pointing from regions of high $\phi$ (red areas) to regions of smaller $\phi$ (yellow areas). Parameters for panels a),b) and c) are $A_{11}=A_{22}=-20$, $A_{12}=-10$, $T=\alpha_T=1$ and $\alpha_A=0.5$. d) Heatmap of $\sigma$ as a function of the dissipative parameters $\alpha_A$ and $\alpha_T$. Numbers written over the heatmap indicate the accuracy $\pi(\overline{\sigma}) = |\sigma-\overline{\sigma}|/\Delta\overline{\sigma}$ of our estimates $\overline{\sigma}$. }
	\label{FIG1}
\end{figure*}

\tcr{As a first example, we consider the Ornstein–Uhlenbeck (OU) process~\cite{weiss2007fluctuation},
\begin{equation}\label{multi_OU}
\dot{\pmb{x}}_t = \pmb{A}\pmb{x}_t + \sqrt{2\pmb{D}}\,\pmb{\xi}_t ,
\end{equation}
where $\pmb{D} = k_{\rm B}\pmb{T}\pmb{\mu}$ and $\pmb{A}$ is the drift matrix. A general discussion of this model is provided in the End Matter; here we focus on the two-dimensional realization of Eq.~\eqref{multi_OU} to test the predictive power of Eq.~\eqref{main_eq}. 
This system, often referred to as the Brownian gyrator~\cite{br_gyr_Villa,br_gyr_2,br_gyr_3,br_gyr_4,loos2020irreversibility,busiello2024unraveling}, is defined by the drift and diffusion matrices
\begin{align}\label{matrices_2D}
\pmb{A} &= 
\begin{pmatrix}
A_{11} & A_{12} \\
\alpha_A A_{12} & A_{22}
\end{pmatrix}, \qquad
\pmb{D} = 
\begin{pmatrix}
T & 0 \\
0 & \alpha_T T
\end{pmatrix},
\end{align}
where, for simplicity, we set $\pmb{\mu} = \mathds{1}$ and $k_{\rm B} = 1$.
} The entropy production rate $\sigma$ can be calculated analytically (see Section S6B in~\cite{SuppMat}) and reads:
\begin{equation}\label{sigma_lin}
    \sigma = \frac{A_{12}^2(\alpha_A - \alpha_T)^2}{\alpha_T\,\inflow} \, .
\end{equation}
From Eq.~\eqref{sigma_lin}, one sees that $\alpha_A$ and $\alpha_T$ are directly related to dissipation. When these parameters differ, mechanical forces and heat fluxes are not balanced, resulting in the generation of probability currents. The magnitude of these currents scales with $A_{12}^2/\alpha_T$ and is modulated by the inflow rate, $\inflow = -(A_{11} + A_{22})$. We thus generated simulated data by fixing $\inflow = 40$ and $A_{12} = 10$, and considered 25 different combinations of the dissipative parameters $\alpha_A$ and $\alpha_T$. For each pair, we simulated a single long trajectory (see Fig.~\ref{FIG1}a for an example) compatible with typical experimental recordings. From these, we derived correlation functions (Fig.~\ref{FIG1}b), as well as $p(\pmb{x}_t)$ and the effective forces $-\nabla\phi(\pmb{x}_t)$ (Fig.~\ref{FIG1}c), {which were used to estimate the entropy production from Eq.~\eqref{main_eq} as $\sigma_{\rm est} = \overline{\sigma} \pm \Delta\overline{\sigma}$, where $\Delta\overline{\sigma}$ denotes the statistical uncertainty. Briefly, a single trajectory is split into subtrajectories, the inference is applied independently, and the results are combined via a weighted average. Technical details are provided in Section~S11 of~\cite{SuppMat}. We emphasize that the analysis pipeline involves specific methodological choices and is not unique. }\tcr{We have also tested the robustness of our approach to measurement noise, a key practical consideration when performing inference from experimental data~\cite{frishman2020learning,ferretti2020building,bruckner2020inferring}, as detailed in Section~S12 of~\cite{SuppMat}.
}

For each estimate $\sigma_{\rm est}$ we assess the quality of the prediction by evaluating the accuracy coefficient $\pi(\overline{\sigma}) = |\sigma-\overline{\sigma}|/\Delta\overline{\sigma} $
that quantifies how far, in units of  statistical error $\Delta\overline{\sigma}$, the estimate $\overline{\sigma}$ is from the true value $\sigma$. The latter is evaluated from the analytical expression \eqref{sigma_lin}. Fig.~\ref{FIG1}d shows the accuracy $\pi$ obtained for all simulated traces, while $\sigma$ is shown in the underlying heatmap. {The inference process yields reliable results: the estimates are well-aligned with the true values, with only moderate deviations relative to their statistical uncertainty. Across all traces, the average accuracy is $\langle \pi \rangle \approx 0.93$, indicating consistent performance of the method within estimated error bounds.}


\begin{figure*}[t!]
 	\centering
\includegraphics[width=1\textwidth]{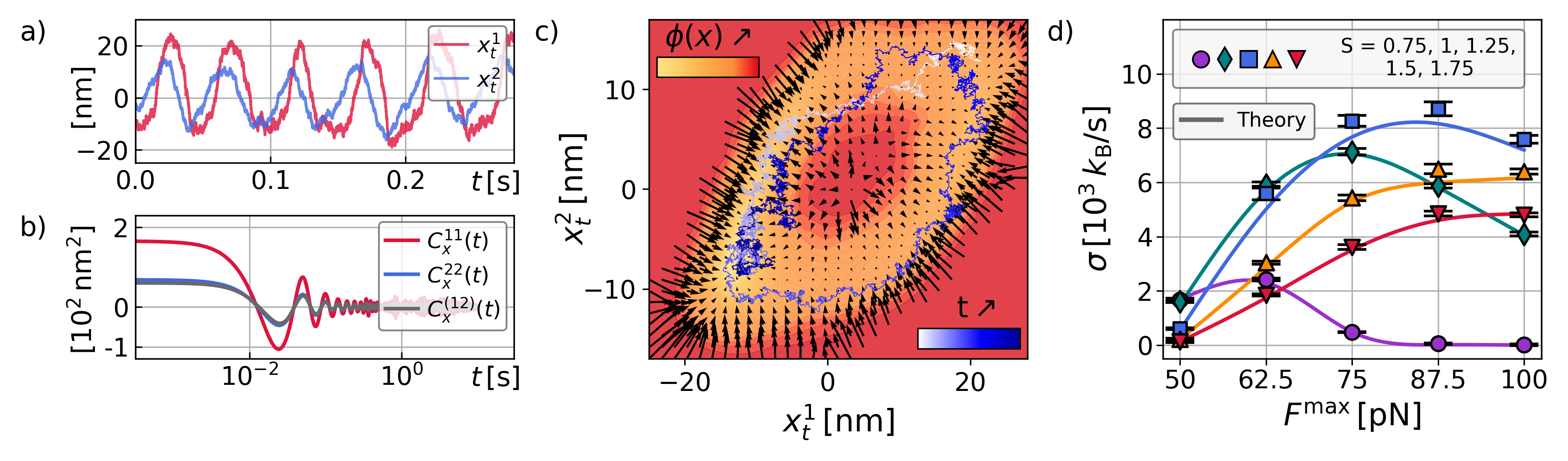}
	\caption{ a) Time series of $x^1_t$ and $x^2_t$ for the hair-bundle model, recorded at a sampling rate of $100\,{\rm kHz}$ over a 10~s interval.
    b) Autocorrelation and cross-correlation functions computed for both degrees of freedom (DOFs). c) Effective forces $-\nabla \phi$ (black arrows) pointing from regions of higher $\phi$ (orange areas) to regions of smaller $\phi$ (yellow areas). The area where the PDF $p(\pmb{x}_t)$ is very small (and $\phi$ is very high) has been highlighted in red. Effective forces have not been shown in this area for visualization purposes. A section of the stochastic trajectory is also depicted on top of the effective potential contour plot. For panels a),b) and c), $F^{\rm max}=100$ and $S=1$. d) Relationship between $\sigma$ and $F^{\rm max}$ for different values of $S$ ($S = 0.75, 1, 1.25, 1.5, 1.75$). Solid lines represent the true $\sigma$ while markers indicate estimated values with associated uncertainties. Note how the estimation procedure demonstrates high reliability across several orders of magnitude of $\sigma$.}
	\label{FIG2}
\end{figure*}

\tcr{As a second example, we examine a nonlinear model of spontaneous hair-bundle oscillations in bullfrog ears~\cite{hair0}, where estimating the entropy production from experimental traces remains an open problem~\cite{roldan2021quantifying,Tucci22}. The model captures the interplay between mechanosensitive ion channels, molecular motors, and calcium feedback through two coupled degrees of freedom, namely the bundle position $x^1$ and the motor position $x^2$, governed by
\begin{equation} \label{eq:SDE_bull}
\begin{split}
\dot{x}^1_t &= -\mu_1 \partial_{x^1}V_t + \sqrt{2k_{\rm B}T\mu_1}\,\xi^1_t ,\\[3pt]
\dot{x}^2_t &= -\mu_2 \partial_{x^2}V_t - \mu_2 F_t^{\rm act} + \sqrt{2k_{\rm B}T^{\rm eff}\mu_2}\,\xi^2_t ,
\end{split}
\end{equation}
where $V_t = V(x^1_t, x^2_t) = V_{\rm E}(x^1_t, x^2_t) + V_{\rm G}(x^1_t, x^2_t) $ describes the mechanical interactions within the system, including elastic forces ($V_{\rm E}$) and the gating dynamics of mechanosensitive ion channels ($V_{\rm G}$) ~\cite{hair1,hair2,hair3}. Active, nonequilibrium driving arises through
\begin{equation}\label{active_bundle}
F_t^{\rm act} = F^{\rm max}\!\left(1 - S\,P_0(x^1_t, x^2_t)\right),
\end{equation}
with $F^{\rm max}$ the maximal motor force and $S$ the calcium-feedback strength. 
The gating probability $P_0(x^1, x^2)$ follows a two-state equilibrium model whose nonlinear dependence on bundle displacement mediates feedback between channel opening and active motor activity~\cite{hair4}. 
Further details are provided in the End Matter.}
While $T^{\rm eff}\approx 1.5 T$ characterizes enhanced fluctuations due to active processes, the main contributors to entropy production are the activity parameters $F^{\rm max}$ and $S$. Following the approach in \cite{roldan2021quantifying}, we focus on the effect of $F^{\rm max}$ and $S$ on $\sigma$, as they directly modulate the nonequilibrium driving forces. To explore a wide range of $\sigma$ values ($1\, k_{\rm B}{\rm/s}$ to $10^4\, k_{\rm B}{\rm/s}$), we simulated 25 traces with varying $F^{\rm max}$ and $S$, keeping all other parameters fixed as in \cite{roldan2021quantifying}. {Trace lengths (10\,s) and sampling rate ($10^5$\,Hz) match typical experimental conditions. We then applied the same subtrajectory-based procedure as in the linear system to compute estimates and uncertainties, using correlation functions and their $t=0$ derivatives for $\traffic$, and kernel density estimation with numerical differentiation for $\inflow$ (Fig.~\ref{FIG2}c).}

As shown in Fig.~\ref{FIG2}d, the predicted values $\overline{\sigma}$ (symbols) closely match the true values $\sigma$ (solid lines). Since the system is nonlinear, the latter were numerically computed using Eq.\eqref{sekimoto}. The accuracy of these estimates can be validated by the relative error, defined as $\delta(\overline{\sigma}) = |\sigma-\overline{\sigma}|/\sigma$. In all cases, the average error remains low at $\langle \delta \rangle \approx 0.05 $, except for instances where $\sigma < 20\, k_{\rm B}{\rm/s}$, where small denominators distort the error measure. The quality of these estimates is further supported by the accuracy metric $\pi(\overline{\sigma})=|\sigma-\overline{\sigma}|/\Delta\overline{\sigma}$, { with  $\langle \pi \rangle \approx 1.67$ and shown in more detail in Fig.~S6a in \cite{SuppMat}}. Crucially, the quality of our estimates remains unaffected by the average ion channel opening probability $\langle P_0 \rangle$ (Fig.~S6b). In contrast, approaches based on trace irreversibility may encounter challenges when $\langle P_0 \rangle \approx 0.5$ \cite{roldan2021quantifying}.
Another key observation is that, similar to the linear model, $\inflow$ remains relatively constant ($\inflow \approx 100\, k_{\rm B}{\rm/s}$ to $500\, k_{\rm B}{\rm/s}$, see Fig.~S6d in \cite{SuppMat}) 
despite large variations in $F^{\rm max}$ and $S$, while $\sigma$ spans orders of magnitude ($1\,k_{\rm B}{\rm/s}$  to $10^4\, k_{\rm B}{\rm/s}$). This indicates that even in this highly nonlinear system, $\traffic$ captures most of the information about the dissipative components of the dynamics.

Our approach can be extended to the cases of partial observations. {This is particularly relevant since, in most systems, only a subset of dynamical degrees of freedom (DOFs), denoted by $\mathcal{S}$, is accessible. In such cases, we can use an intermediate result of Eq.~\eqref{main_eq} (see Section S1 in~\cite{SuppMat}):
\begin{equation}
    -\ddotnice{C}^{\, ii}_{x}(0) \le C_{\nu}^{\,ii}(0) = \langle (\nu^i)^2 \rangle\,,
\end{equation}
where $\nu^i$ is the $i^{\rm\, th}$ component of the mean local velocity \eqref{eq:Anu_main}. For any $i \in \mathcal{S}$, if $\ddotnice{C}^{\, ii}_{x}(0) < 0$, it implies $C_{\nu}^{\,ii}(0) > 0$, therefore implying a non-equilibrium state. This provides a practical criterion for detecting nonequilibrium under partial observation. For a one dimensional Gaussian trace this is particularly challenging ~\cite{lucente2022inference}: time reversal tests fail ~\cite{zamponi2005fluctuation,caprini2019entropy}, and in one dimension probability currents are not observable unless the topology is nontrivial, which rules out current based methods such as the TUR. When $\pmb{D}=k_{\rm B}\pmb{T}\pmb{\mu}$, our approach offers a complementary route to characterize nonequilibrium when irreversibility based methods are insufficient. }

{When the diffusion matrix is diagonal, Eq.~\eqref{main_eq} not only enables non-equilibrium detection but also provides an effective lower bound on $\sigma$. This is particularly relevant in systems like the hair-bundle model \eqref{eq:SDE_bull}, where direct observation of certain degrees of freedom, such as the motor coordinate $x^2_t$, is experimentally challenging. Moreover, having prior knowledge that the system can be effectively described by a model with a diagonal diffusion matrix facilitates the practical application of this bound. More in detail,} as shown in Section S4 in \cite{SuppMat}, 
it is useful to decompose $\sigma$ as
\begin{equation}
\sigma = \sum_i \sigma_{i} = \sum_i \left(4 \traffic_i + \inflow_i\right)\,. 
\end{equation}
Each component $\sigma_i$ is defined by $\sigma_i = \langle(\nu^i)^2\rangle / D^{ii} \ge 0$, $\traffic_i = -\,\ddot{C}_{x}^{\,ii}(0)\,\bigl/\,\bigl(4\,D^{ii}\bigr)$, and $\inflow_i = D^{ii}\,C_{\nabla \phi}^{\,ii}(0) = D^{ii}\mathrm{Var}\bigl(\partial_{\,i}\phi\bigr) \ge 0$. From this, it follows
\begin{equation}\label{ent_bound}
4 \traffic_i = \sigma_i - \inflow_i \leq \sigma_i \leq \sigma \, ,
\end{equation}
implying that any positive traffic component $\traffic_i > 0$ guarantees $\sigma > 0$, indicating the system is out of equilibrium and providing a lower bound to the entropy production rate. This observation thus provides a practical method for detecting nonequilibrium and bounding $\sigma$ potentially from a single stochastic trace. More in general, for a subset $\mathcal{S}$ of observed DOFs, one can select only positive traffic components and obtain a partial estimate of $\sigma$, namely 
\begin{equation} \label{sigma_part}
    \sigma_{\cal S} = \sum_{i \in \mathcal{S}} {\rm max}(4 \traffic_{i}\, ,0) \le \sigma \, .
\end{equation}
Similar kinetic bounds on $\sigma$ have been previously reported in \cite{maes2017frenetic}, though Eq.~\eqref{sigma_part} has the advantage of being experimentally accessible. We note that estimating $\inflow$ requires knowledge of the PDF of the entire system, making it impossible to derive from partial observations alone.
\begin{figure}[t]
 	\centering \includegraphics[width=0.8\linewidth]{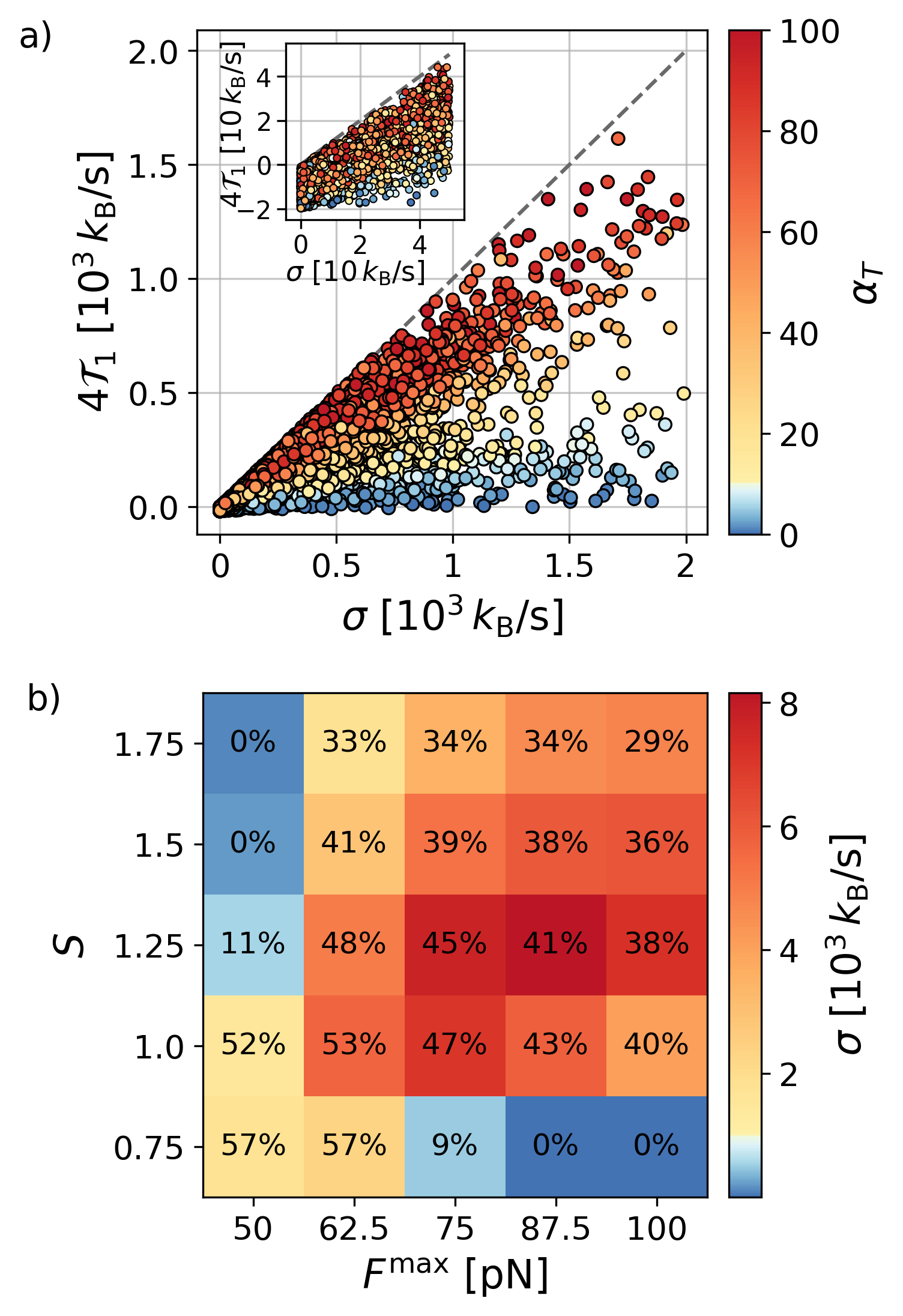}
\caption{a) Scatter plot of $4\mathcal{T}_1$ vs. $\sigma$, illustrating the lower bound \eqref{ent_bound} for randomly sampled parameters: $A_{11} \in [-20, -1]$, $A_{22} \in [-20, -1]$, $A_{12} \in [-10, 0]$, $\alpha_A \in [-50, 1]$, and $\alpha_T \in [0.1, 100]$. Color bar indicates $\alpha_T$, showing how increasing thermal bath asymmetry tightens the bound. b) Heatmap of $\sigma$ as a function of $F^{\rm max}$ and $S$. Numerical values in the heatmap correspond to the fraction $4{\cal T}_1/\sigma$, representing the proportion of $\sigma$ estimated by the bound.}
	\label{FIG3}
\end{figure}
The Gaussian system defined by \eqref{matrices_2D}, despite its simplicity, offers valuable insights when applying Eq.~\eqref{ent_bound} under the minimal assumption of a diagonal diffusion matrix. Fig.~\ref{FIG3}a illustrates the relation between $4\traffic_1$ and $\sigma$ for various parameters in \eqref{matrices_2D}, demonstrating that a higher ratio between the temperatures of the two thermal baths ($\alpha_T$) tightens the bound, allowing for more precise nonequilibrium detection. Furthermore, the condition $\traffic_1 > 0$ corresponds to the absence of an equilibrium solution for the 2D linear system leading to the observed correlation function $C^{\, 11}_{x}(t)$ (see Section S8 in \cite{SuppMat} 
for details). In our previous work~\cite{VSR}, red blood cell (RBC) flickering traces were Gaussian and yielded $\traffic_{1}<0$ for the observed degree of freedom, requiring a specific model for thermodynamic characterization. Section~S9 of~\cite{SuppMat} analyzes the detectability of $\sigma$ under partial observation in linear systems, including the RBC case from~\cite{VSR}.


For the model \eqref{eq:SDE_bull}, our bound yields the results shown in Fig.~\ref{FIG3}b. The colors of the heat map represent $\sigma$, while the numbers indicate the fraction of $\sigma$ estimated using our method, specifically $4\traffic_1 / \sigma$. In the more active regions of parameter space, the estimator recovers roughly half of $\sigma$, yielding inferred dissipation about three orders of magnitude larger than the values reported in \cite{roldan2021quantifying,ghosal2022inferring}. This shows that the method is able to capture a substantial fraction of entropy production and is well suited to study biological systems under partial observation.

\tcr{To conclude, this Letter introduces Eq.~\eqref{main_eq} to estimate the entropy production rate $\sigma$ directly from stochastic trajectories without forces or steady-state currents. Unlike current-based approaches, we infer $\sigma$ from kinetic information, such as the short-time curvature (traffic) of correlation functions. This is important for experiments that record only positions.
Other practical advantages are that (i)
our approach applies under partial observation (we find evidence of tight bounds in cases where the TUR performs poorly, as discussed in Section S9B in \cite{SuppMat}), and (ii) even applies when only a single degree of freedom is observed, where currents are not detectable. We illustrate this both in Gaussian systems coupled to multiple thermal baths, where detecting and quantifying nonequilibrium from a single degree of freedom is notoriously challenging~\cite{zamponi2005fluctuation,caprini2019entropy,lucente2022inference}, and in a biological model of hair cells, where we infer dissipation from a single measured coordinate (Fig.~\ref{FIG3}b). We also provide practical procedures to estimate the traffic from discrete data (Section S12 in \cite{SuppMat}) and demonstrate robustness to measurement noise (Section S13 in \cite{SuppMat}).  
Overall, our approach offers a powerful and versatile framework for analyzing dissipation in nonequilibrium diffusive systems, including single particle tracking and microrheology \cite{Opt_tweez_review}, active matter systems \cite{foster2023dissipation}, and for biological settings with partial observation \cite{hair0}.}




\begin{acknowledgments}
\begin{center}
    {\bf ACKNOWLEDGMENTS}
\end{center}
The author is especially grateful to Marco Baiesi for his insightful input, and thanks Felix Ritort, Vincenzo Maria Schimmenti, Matteo Ciarchi, Daniel Maria Busiello, and Edgar Roldan for valuable discussions and feedback. 
\end{acknowledgments}


\bibliography{Bibliography}

\clearpage

\onecolumngrid
\begin{center}
\vspace{0.5cm}
\textbf{\large End Matter}\\
\end{center}
\twocolumngrid

\tcr{\textit{Multivariate Ornstein–Uhlenbeck model} -We consider the multivariate Ornstein–Uhlenbeck process~\cite{weiss2007fluctuation},
\begin{equation}\label{multi_OU_supp}
\dot{\pmb{x}}_t = \pmb{A}\pmb{x}_t + \sqrt{2\pmb{D}}\,\pmb{\xi}_t ,
\end{equation}
where $\pmb{x}_t \in \mathbb{R}^n$ is the state vector, $\pmb{A}$ is the drift matrix, $\pmb{D}$ the diffusion matrix, and $\pmb{\xi}_t$ a vector of independent Gaussian white noises with $\langle \xi^i_t\xi^j_{t'} \rangle = \delta^{ij}\delta(t-t')$.  
The process is dynamically stable provided all eigenvalues of $\pmb{A}$ have negative real parts, $\Re(\lambda_i) < 0$ for all $i$. The diffusion matrix relates to mobility and temperature as
\begin{equation}
\pmb{D} = k_{\rm B}\pmb{T}\pmb{\mu},
\end{equation}
where $\pmb{\mu}$ is the mobility tensor and $\pmb{T}$ may encode local effective temperatures associated with different thermal reservoirs.  
The drift matrix can be written as
\begin{equation}\label{eq:drift_decomp}
\pmb{A} = \pmb{\mu}\left(-\pmb{K} + \pmb{\psi}\right),
\end{equation}
where $\pmb{K}$ is a symmetric positive-definite matrix and $\pmb{\psi}$ is an antisymmetric matrix.   
The symmetric matrix $\pmb{K}$ may be associated with an effective quadratic potential, while $\pmb{\psi}$ may represent non conservative forces.  
No particular physical meaning needs to be ascribed to $\pmb{K}$ and $\pmb{\psi}$ unless additional structure is specified for $\pmb{\mu}$.}
  
\tcr{Nonequilibrium conditions arise either from asymmetries in $\pmb{\psi}$ or from temperature heterogeneity encoded in $\pmb{T}$.  
The inflow rate, defined as the negative trace of the drift matrix, reads
\begin{equation}
\inflow = -\mathrm{Tr}(\pmb{A}) = \mathrm{Tr}(\pmb{\mu}\pmb{K})
\end{equation}
and depends only on the symmetric part of the drift, because $\mathrm{Tr}(\pmb{\mu}\pmb{\psi}) = 0$ when $\pmb{\mu}$ is symmetric and $\pmb{\psi}$ antisymmetric.  
Consequently, the inflow is insensitive to the sources of nonequilibrium, such as rotational or non-reciprocal couplings and temperature differences.  
This property extends to the case where $\pmb{\psi}$ is non-symmetric but has vanishing diagonal elements, provided that $\pmb{\mu}$ (and thus $\pmb{D}$) is diagonal, since any diagonal elements in $\pmb{\psi}$ can be absorbed into $\pmb{K}$ without loss of generality (see Section S6 of~\cite{SuppMat} for more details). 
When $\pmb{\mu}$ is diagonal, the matrix $\pmb{K}$ can be associated to conservative forces arising from a quadratic potential while the non-symmetric $\pmb{\psi}$ corresponds to non-reciprocal forces that drive the system out of equilibrium. For this case, and from Eq.~\eqref{main_eq_exp}, it also follows that in this for a fixed potential, increases in entropy production $\sigma$ due to $\pmb{\psi}$ must captured by the traffic $\traffic$. }

\tcr{\textit{Hair-bundle model} - The spontaneous oscillations of hair bundles in the auditory organs of bullfrogs are driven by the interplay of mechanosensitive ion channels, molecular motor activity, and calcium feedback mechanisms~\cite{hair0,hair1,hair2,hair3,roldan2021quantifying}. 
These dynamics are effectively captured by a nonlinear model with two coupled degrees of freedom: the bundle position $x^1$ and the center of mass of the molecular motors $x^2$.}

\tcr{The stochastic equations of motion read
\begin{equation}
\begin{split}
\dot{x}^1_t &= -\mu_1 \partial_{x^1}V_t + \sqrt{2k_{\rm B}T\mu_1}\,\xi^1_t ,\\[3pt]
\dot{x}^2_t &= -\mu_2 \partial_{x^2}V_t - \mu_2 F_t^{\rm act} + \sqrt{2k_{\rm B}T^{\rm eff}\mu_2}\,\xi^2_t ,
\end{split}
\end{equation}
where $\mu_1$ and $\mu_2$ are mobility coefficients and $V_t = V(x^1_t, x^2_t)$ is the total potential energy. 
The potential includes an elastic ($V_{\rm E}(x^1, x^2)$) contribution and a gating ($V_{\rm G}(x^1, x^2)$) contribution,
\begin{align}
V_{\rm E}(x^1, x^2) &= \frac{k_{gs}}{2}(x^1 - x^2)^2 + \frac{k_{sp}}{2}(x^1)^2 ,\\[3pt]
V_{\rm G}(x^1, x^2) &= -N k_{\rm B}T \ln\!\left[\exp\!\left(\frac{k_{gs}D(x^1 - x^2)}{N k_{\rm B}T}\right) + A\right],
\end{align}
so that $V(x^1, x^2) = V_{\rm E}(x^1, x^2) + V_{\rm G}(x^1, x^2)$. 
Here $k_{gs}$ and $k_{sp}$ are stiffness coefficients, $D$ is the gating swing of a transduction channel, and 
\begin{equation}
A = \exp\!\left[\frac{\Delta G + k_{gs}D^2/(2N)}{k_{\rm B}T}\right],
\end{equation}
where $\Delta G$ denotes the free-energy difference between open and closed channel states, and $N$ is the number of transduction elements.}

\tcr{The system is driven out of equilibrium by molecular motor activity, represented by an effective temperature $T^{\rm eff}$ and a non-conservative active force
\begin{equation}
F_t^{\rm act} = F^{\rm max}\!\left(1 - S\,P_0(x^1_t, x^2_t)\right),
\end{equation}
where $F^{\rm max}$ is the maximal motor force and $S$ quantifies the strength of calcium-mediated feedback on the motors. 
The gating probability of the mechanosensitive ion channels is
\begin{equation}
P_0(x^1, x^2) = \frac{1}{1 + A \exp\!\left[-\frac{k_{gs}D(x^1 - x^2)}{N k_{\rm B}T}\right]},
\end{equation}
which represents the open probability of a two-state equilibrium model of channel gating. 
In this model, the free-energy difference between open and closed states depends linearly on the displacement $x^1 - x^2$. 
In practice, $P_0(x^1, x^2)$ couples channel gating to bundle motion, mediating feedback between mechanical displacement and active motor forces. 
This feedback loop underlies the self-sustained oscillations characteristic of active hair-bundle dynamics~\cite{hair4}.}

\clearpage
\setcounter{equation}{1}
\setcounter{figure}{0}
\setcounter{table}{0}
\setcounter{page}{0}
\makeatletter
\renewcommand{\theequation}{S\arabic{equation}}
\renewcommand{\thefigure}{S\arabic{figure}}
\setcounter{secnumdepth}{2} 
\renewcommand{\thesection}{S\arabic{section}} 

\onecolumngrid

\begin{center}
{\LARGE Supplemental Material for}\\
\vspace{0.4cm}
\textbf{\Large Force-Free Kinetic Inference of Entropy Production}\\
\hspace{0.5cm}

{I. Di Terlizzi}

\hspace{0.5cm}
{Max Planck Institute for the Physics of Complex Systems, N{\"o}thnitzer Stra{\ss}e 38, 01187, Dresden, Germany}

\end{center}

\section{Main result derivation}\label{main_derivation}

This section is devoted to the derivation of the main result \eqref{main_eq}. As a starting point, we consider the Langevin equations
\begin{equation} \label{LE_supp}
\dot{x}^i_t = A^i_t + \sqrt{2D^{ij}}\xi^j_t \, ,
\end{equation}
where, using Einstein's summation convention, we define the drift $A^i_t=A^i(\pmb{x}_t)=\mu_{ij} F^j_t$ for simplicity and the Gaussian noise $\xi^i_t$ has first and second moments given by
$\langle \xi^i_t \rangle = 0$ and $\langle \xi^i_t\, \xi^j_s \rangle = \delta^{ij} \delta(t-s)$. The associated Fokker-Planck equation is given by
\begin{equation}\label{FP_supp}
\partial_t p(\pmb{x}_t) = - \nabla \cdot \pmb{j}(\pmb{x}_t),
\end{equation}
where $\pmb{j}(\pmb{x}_t) = \pmb{\nu}(\pmb{x}_t)p(\pmb{x}_t) = \left(\pmb{A}(\pmb{x}_t) - \nabla \phi(\pmb{x}_t)\right)p(\pmb{x}_t)$ represents the \textit{probability current}, and $\pmb{\nu}(\pmb{x}_t)p(\pmb{x}_t)$ is the \textit{mean local velocity}. At equilibrium, the probability current vanishes, $\pmb{j}(\pmb{x}_t) = 0$, leading to a stationary probability density function (PDF) where $\partial_t p(\pmb{x}_t) = 0$. In contrast, a nonequilibrium steady state (NESS) is characterized by non-zero, irrotational probability currents satisfying $\nabla \cdot \pmb{j}(\pmb{x}_t) = 0$, which also results in a time-independent PDF.

The variance sum rule \cite{VSR, VSR_solved_models} for the displacement $\Delta x^i_t = x^i_t-x^i_0$ and the drift $A^i_t$ 
is
\begin{align}
\label{eq:VSR1}
    \frac 1 2 \Cov{\Delta x^i_t}{\Delta x^j_t} 
    + \int_0^t \dd t^{\prime} \int_0^{t'} \dd t^{\prime\prime} ~\Cov{\AA^i_{t^{\prime\prime}}}
    {\AA^j_0}
    =
    D^{ij} t + 2 \int_0^t \dd t^{\prime} \int_0^{t^{\prime}} \dd t^{\prime\prime} \Sym{ \Cov{\dot x^i_{t^{\prime\prime}}}{\nu^j_0} }
\end{align}
where $\Sym{\cdot}$ denotes the symmetrisation of indexes $i$ and $j$. As shown in \cite{VSR_solved_models}, this formula holds for a diffusion matrix $ \pmb{D} $ that can be expressed in either of the following forms: $ \pmb{D} = k_{\rm B} T \pmb{\mu} $, where $ T $ is the temperature and $ \pmb{\mu} $ is an arbitrary positive-definite symmetric mobility matrix, or $ \pmb{D} = k_{\rm B} \pmb{T} \pmb{\mu} $, where $ \pmb{T} $ is a diagonal matrix of temperatures and $ \pmb{\mu} $ is a diagonal mobility matrix. From a practical point of view, the drifts $\AA^i$ are usually be difficult to measure in experiments. Hence, we aim to replace them with functions of the effective potential
\begin{align}
    \phi(\pmb{x}_t) = -\log p(\pmb{x}_t) +C
\end{align}
which, we assume, can be estimated from recorded trajectories and with C a constant depending on the normalization of $p(\pmb{x}_t)$. 
In particular, we consider the effective forces introduced above as the gradient of $\log p(\pmb{x})$,
\begin{align}
    g^i(\pmb{x}_t) = -\partial_i \phi(\pmb{x}_t)
\end{align}
and the related "Fick" velocities
\begin{align}
\label{eq:u}
    u^i(\pmb{x}_t) = D^{ij} g^j(\pmb{x}_t) 
\end{align}
In this way, the drift
\begin{align}
    \AA^i(\pmb{x}_t) = \nu^i(\pmb{x}_t) + u^i(\pmb{x}_t)
    \label{eq:Anu}
\end{align}
is interpreted as a sum of the local mean velocity $\vec \nu$ and the Fick velocity $\vec u$.
The VSR is rewritten as
\begin{align}
\label{eq:VSR2}
    \frac 1 2 \Cov{\Delta x^i_t}{\Delta x^j_t}
    =
    D^{ij} t +
    \int_0^t \dd t' \int_0^{t'} \dd t^{\prime\prime}
    \Sym{ 2\,\Cov{\dot x^i_{t^{\prime\prime}}}{\nu^j_0} 
    -\Cov{\AA^i_{t^{\prime\prime}}}{\AA^j_0}}
\end{align}
and then use \eqref{LE_supp} to replace $\dot x^i_t$ and then \eqref{eq:Anu} to replace the drift $\AA^i_t$.
Thus, the term in the integral becomes
\begin{align}
    & 2 \Sym{ \Cov{\dot x^i_t}{\nu^j_0} }
    - \Sym{ \Cov{\AA^i_t}{\AA^j_0} }
    = \nonumber\\
    &\quad=2 \Sym{\Cov{\AA^i_t}{\nu^j_0} + \underbrace{\Cov{\sqrt{2 D}^{ik} \xi^k_t}{\nu^j_0}}_{0} }
    - \Sym{ \Cov{\AA^i_t}{\AA^j_0} }
    \nonumber\\
    &\quad=
    2 \Sym{\Cov{\nu^i_t}{\nu^j_0} +\Cov{u^i_t}{\nu^j_0} } 
    \nonumber\\
    & \qquad- \Sym{ \Cov{\nu^i_t}{\nu^j_0} + \Cov{u^i_t}{u^j_0} + \Cov{\nu^i_t}{u^j_0} + \Cov{u^i_t}{\nu^j_0} }
    \nonumber\\
    &\quad=
    \Sym{ \Cov{\nu^i_t}{\nu^j_0} + \Cov{u^i_t}{\nu^j_0}
    - \Cov{u^i_t}{u^j_0} - \Cov{\nu^i_t}{u^j_0}  }
\end{align}
and we may rewrite the VSR as
\begin{align}
\label{eq:VSR3}
    \frac 1 2 \Cov{\Delta x^i_t}{\Delta x^j_t}
    + & \int_0^t \dd t' \int_0^{t'} \dd t^{\prime\prime} \Sym{\Cov{u^i_{t^{\prime\prime}}}{u^j_0}}=
    \nonumber\\
    & =
    D^{ij} t + 
    \int_0^t \dd t' \int_0^{t'} \dd t^{\prime\prime} \Sym{ \Cov{\nu^i_{t^{\prime\prime}}}{\nu^j_0} + \Cov{u^i_{t^{\prime\prime}}}{\nu^j_0} - \Cov{\nu^i_{t^{\prime\prime}}}{u^j_0} }
\end{align}
The second time derivative of \eqref{eq:VSR3}, evaluated at time $t=0$,
\begin{align}
\label{eq:d_VSR3}
    \frac 1 2 \!\left.\partial^2_t\Cov{\Delta x^i_t}{\Delta x^j_t}\right|_{t=0} + \Cov{u^i}{u^j}
    =
    \Cov{\nu^i}{\nu^j} + 
    \underbrace{\Sym{\Cov{u^i}{\nu^j} - \Cov{\nu^i}{u^j} }}_{0}\,.
\end{align}
contains a term that we may re-cast as $\Cov{\nu^i}{\nu^j} = \smean{\nu^i\nu^j} $ because $v^i=\smean{\nu^i}=0$ in a NESS. Introducing $\smean{\nu^i\nu^j}$ is convenient because the entropy production rate is the weighted sum of such terms,
\begin{align}
 \label{eq:sigma0}
     \sigma = (D^{-1})^{ij} \mean{\nu^i\nu^j} \, .
\end{align}
Hence, using~\eqref{eq:d_VSR3}, we rewrite~\eqref{eq:sigma0} as 
\begin{align}
\label{eq:sigma1}
\sigma = (D^{-1})^{ij} 
\left[\frac 1 2 \!\left.\partial^2_t\Cov{\Delta x^i_t}{\Delta x^j_t}\right|_{t=0}  + \Cov{u^i}{u^j}
\right]
\end{align}
By using that, in a NESS,
\begin{align}
    \frac 1 2 \!\left.\partial^2_t\Cov{\Delta x^i_t}{\Delta x^j_t}\right|_{t=0} = \partial^2_t \!\left( C^{\, ij}_x(0)^{ij} - C^{\, ij}_x(t) \right)|_{t=0}=-\ddotnice{C}^{\,(ij)}_x(0)\, ,
\end{align}
where $(ij)$ denotes symmetrised indexes, along with~\eqref{eq:u}, we may also rewrite~\eqref{eq:sigma1} as 
\begin{align}
\label{eq:sigma2}
\sigma 
&=   - (D^{\rm -1})^{ij} 
 \ddotnice{C}_{x}^{\, (ij)}(0)
 + D^{ij}\Cov{g^i}{g^j}  =  {\rm Tr}\!\left[-\pmb{D}^{-1} \,\ddotnice{\pmb{C}}^{\, \rm S}_x(0)+\pmb{D}\pmb{C}_{\nabla\phi}(0)\right]  =  4\traffic + \inflow 
\end{align}
to highlight the presence of the traffic $\traffic$ and inflow rate $\inflow$ in the formula. Note that ${\rm Tr}[\cdot]$ is the trace operator, $\pmb{C}^{\, \rm S}_{O}(t)=(\pmb{C}_{O}(t)+\pmb{C}^{\, \rm T}_{O}(t))/2$ denotes the symmetrized matrix, and that, since $\pmb{D}^{\rm-1}$ is symmetric, it follows that $\traffic = {\rm Tr}[\pmb{D}^{-1} \,\ddotnice{\pmb{C}}^{\, \rm S}_x(0)] = {\rm Tr}[\pmb{D}^{-1}\, \ddotnice{\pmb{C}}_x(0)]$, as the contraction of the antisymmetric part of a tensor with a symmetric tensor is zero. 

\section{\texorpdfstring{Proof that $\traffic = -{\rm Tr}\!\left[\pmb{D}^{-1} \ddotnice{\pmb{C}}^{\, \rm S}_{x}(0) \right]$}{Proof that T = T}}
\label{proof_T}

We start from the definition of traffic presented in \cite{maes2008steady}:
\begin{equation}\label{traffic}
    \traffic = \frac{1}{4}\int \dd \pmb{x}_t \, p(\pmb{x}_t)  A^i_t (D^{\rm -1})^{ij} A^j_t  +\frac{1}{2} \int \dd \pmb{x}_t\, p(\pmb{x}_t) \partial_i  A^i_t \, .
\end{equation}
By plugging $ \AA^i = \nu^i + u^i=  \nu^i - D^{ij}\partial_j \phi_t $ (Eq. \eqref{eq:Anu}) in the first term on the right hand side, and by using the definition of $\sigma$ in \eqref{eq:sigma0}, one gets
\begin{equation}\label{traffic_step1}
    \traffic = \frac{\sigma}{4} + \frac{\inflow}{4} - \frac{1}{2}\int \dd  \pmb{x}_t \Sym{ \, p(\pmb{x}_t) \nu^i_t \partial_i \phi_t }   +\frac{1}{2} \int \dd \pmb{x}_t\, p(\pmb{x}_t) \partial_i  A^i_t \, .
\end{equation}
For the second term on the right-hand side, it follows directly that
\begin{equation}
\int \dd \pmb{x}_t \Sym{ p(\pmb{x}_t)\nu^i_t \partial_i \phi_t } = - \int \dd \pmb{x}_t \Sym{\partial_i \!\left( p(\pmb{x}_t) \nu^i_t \right) \phi_t } = 0 \, ,
\end{equation}
where we have used the fact that $p(\pmb{x}_t)$ decays sufficiently rapidly at infinity and that $\partial_i \! \left( p(\pmb{x}_t) \nu^i_t \right) = 0$ in a NESS, as discussed after Eq. \eqref{FP_supp}. Moreover, by using the definition of $\inflow$
\begin{equation}\label{inflow_for_traffic}
\begin{split}
    \inflow &= \int \dd \pmb{x}_t \, p(\pmb{x}_t)  u^i_t (D^{\rm -1})^{ij} u^j_t  \\
    &= \int \dd \pmb{x}_t \, p(\pmb{x}_t) (A^i_t -\nu_t^i)\, \partial_i \log p(x_t) \\
    &= \int \dd \pmb{x}_t \, A^i_t \, \partial_i p(\pmb{x}_t) + \int \dd \pmb{x}_t \, p(\pmb{x}_t) \nu_t^i \, \partial_i \log p(\pmb{x}_t) =\\
    & = - \int \dd \pmb{x}_t \, p(\pmb{x}_t)\, \partial_i  A^i_t - \int \dd \pmb{x}_t  \log p(\pmb{x}_t) \partial_i \!  \left( p(\pmb{x}_t) \nu_t^i \right)   =- \int \dd \pmb{x}_t \,  p(\pmb{x}_t) \, \partial_i  A^i_t \,
\end{split}
\end{equation}
where again we leveraged the rapid decay of $p(\pmb{x}_t)$ at infinity and the propriety that $\partial_i \! \left( p(\pmb{x}_t) \nu^i_t \right) = 0$ in a NESS. By finally combining \eqref{traffic_step1} and \eqref{inflow_for_traffic} we get 
\begin{equation}
    \traffic = \frac{1}{4}(\sigma - \inflow) \, ,
\end{equation}
which compared to Eq. \eqref{eq:sigma2} concludes the proof.


\section{Scalings in the large dissipation regime}\label{InfoG}

{We consider an overdamped Langevin system at homogeneous temperature $T$, with constant, possibly non-diagonal, mobility matrix $\pmb{\mu}$ and a force field of the form 
\begin{equation}
    \pmb{F}(\pmb{x}_t) = -\nabla V(\pmb{x}_t) + \alpha\, \pmb{B}(\pmb{x}_t)\, .
\end{equation} 
In this setting, the system is driven out of equilibrium by the non-conservative force $\alpha\, \pmb{B}(\pmb{x}_t)$, while the steady-state density $p(\pmb{x}_t)$ is assumed to remain normalized and confined by the potential $V(\pmb{x}_t)$ that grows at infinity.}

{In the large $\alpha$ limit, the mean local velocity becomes dominated by the non-conservative part of the drift:
\begin{equation}
\pmb{\nu}(\pmb{x}_t) = \pmb{\mu} \pmb{F}(\pmb{x}_t) + \pmb{D} \nabla \phi(\pmb{x}_t) \sim \alpha\, \pmb{\mu} \pmb{B}(\pmb{x}_t).
\end{equation}
At leading order in $\alpha$, the entropy production rate hence becomes
\begin{equation}
\sigma = \left\langle \pmb{\nu}^\top \pmb{D}^{-1} \pmb{\nu} \right\rangle \sim \frac{\alpha^2}{T} \left\langle \pmb{B}^\top \pmb{\mu} \pmb{B} \right\rangle.
\end{equation}
Since $p(\pmb{x}_t)$ remains regular and confined, all steady-state averages involving $\pmb{B}(\pmb{x}_t)$ are expected to remain of the same order in $\alpha$, unless $\pmb{B}$ exhibits singularities. For the inflow rate, defined in \eqref{inflow_for_traffic}, we get
\begin{equation}
\inflow = -\left\langle \nabla \cdot (\pmb{\mu} \pmb{F}) \right\rangle = \left\langle \nabla \cdot (\pmb{\mu} \nabla V) \right\rangle - \alpha \left\langle \nabla \cdot (\pmb{\mu} \pmb{B}) \right\rangle \sim - \alpha \left\langle \nabla \cdot (\pmb{\mu} \pmb{B}) \right\rangle,
\end{equation}
since the conservative contribution is $\alpha$-independent. Similarly, the traffic is given by \eqref{traffic}
\begin{equation}\label{traffic_scaling}
\traffic = \frac{1}{4} \left\langle \pmb{A}^\top \pmb{D}^{-1} \pmb{A} \right\rangle + \frac{1}{2} \left\langle \nabla \cdot \pmb{A} \right\rangle, \quad \text{with } \pmb{A} = \pmb{\mu} \pmb{F},
\end{equation}
so that, when $\pmb{A} \sim \alpha\, \pmb{\mu} \pmb{B}$, at leading order we find
\begin{equation}
\traffic \sim \frac{\alpha^2}{4T} \left\langle \pmb{B}^\top \pmb{\mu} \pmb{B} \right\rangle + \frac{\alpha}{2} \left\langle \nabla \cdot (\pmb{\mu} \pmb{B}) \right\rangle.
\end{equation}}

{Hence, both $\sigma$ and $\traffic$ grow quadratically in $\alpha$, while $\inflow$ grows linearly. This confirms that in the large dissipation regime, the entropy production is dominated by traffic, as described by the identity:
\begin{equation}
\sigma = 4\, \traffic + \inflow\,.
\end{equation}}

{Physically, $\inflow$ reflects how concentrated or spread out the steady-state density $p(\pmb{x}_t)$ is in space, depending on how it responds to the divergence of the drift field. In contrast, $\sigma$ quantifies the activity of the system through the squared mean local velocity. Even under strong non-conservative forcing, the density remains confined by the potential and primarily redistributes along streamlines of $\pmb{B}(\pmb{x}_t)$, leading to a relatively slow increase in $\inflow$ compared to the steep quadratic growth of $\sigma$.}

\section{\texorpdfstring{Partial observations and bounds for diagonal $\pmb{D}$}{Partial observations and bounds for diagonal D}}\label{partial_obs}

We start from our main result \eqref{main_eq} for the particular case of a diagonal diffusion matrix $\pmb{D}$. In this case, following the derivation in Section \ref{main_derivation}, one can write $\sigma$ as 
\begin{equation}\label{sigma_decomponition_phi}
    \sigma = \sum_i \sigma_{i} = \sum_{i} \left(4 \traffic_i + \inflow_i\right)
\end{equation}
where the equality holds component-wise with terms given by
\begin{align}\label{sigma_components_phi}
     \sigma_{i} = \frac{\langle (\nu^i)^2  \rangle}{D^{ii}} > 0 && \traffic_i = - \frac{\ddotnice{C}_{x}^{\, ii}(0)}{4 D^{ii}} && \inflow_i = D^{ii} C^{\,ii}_{\nabla \phi} (0) = D^{ii}{\rm Var}\left( \partial_i \phi  \right) \ge 0 \, .
\end{align}
As a first consequence of this component-wise relationship is that 
\begin{equation}
    4\traffic_i = \sigma_{i} -\inflow_i \le \sigma_{i} \le \sigma
\end{equation}
implying that, whenever a traffic component is bigger that zero, it also means that $\sigma$ is non-zero and the system is out of equilibrium. This is a very important feature of \eqref{sigma_decomponition_phi} as it enables the detection of non-equilibrium even from a single stochastic trace and provides a practical way to lower bound the entropy production rate. This result generalises the criterion presented in \cite{VSR_solved_models} for which, if  $\sigma_c = \sum_{i} 2 \traffic_{i}$ is positive, then $\sigma \ge \sigma_c$ is large and the system is strongly out of equilibrium. Indeed, the latter argument, could only be applied if all degrees of freedom can be observed. Here instead, one can rely on limited observations of a subset $\mathcal{S}$ of all DOFs, select only positive traffic components and get a partial estimate of $\sigma$, namely 
\begin{equation}\label{bound_sigma}
    \sigma_{\mathcal{S}} = \sum_{i \in \mathcal{S}} {\rm max}(4 \traffic_{i},0) \le \sigma \, .
\end{equation}
We stress that, in order to estimate $\inflow$ one needs the complete PDF of the whole system and hence, it can not be estimated from partial observations. 

To conclude this section, we compare Eq. \eqref{sigma_decomponition_phi} with the expression for $\sigma$ derived in \cite{VSR,VSR_solved_models}, which is given by
\begin{equation}\label{sigma_decomponition_F}
\sigma = \sum_i \sigma_i^{F} = \sum_i \left(2 \traffic_i + \frac{\mu_i}{2 k_{\rm B} T_i } {\rm Var}\left( F^i_t \right) \right),
\end{equation}
where, as before, the equality holds component-wise. Here, $F_i$, $\mu_i$, and $T_i$ denote the force, mobility, and temperature associated with the $i^{\text{th}}$ degree of freedom (DOF), respectively. In this case, the term $\sigma_i^{F} = \langle F^i_t \circ \dot{x}_t \rangle / k_{\rm B} T$ represents the heat flux transferred to the environment by the $i^{\text{th}}$ DOF. Notably, this component can be either positive or negative depending on the direction of the flow. This implies that measuring microscopic forces allows the determination of DOF-specific heat fluxes, which cannot be achieved by relying solely on traffic and effective forces. However, while Eq. \eqref{sigma_decomponition_F} provides insights into these specific fluxes, it does not allow the derivation of thermodynamic bounds, such as those established in Eq. \eqref{bound_sigma}.

\section{\texorpdfstring{Proof that $\pmb{D} = -\dotnice{\pmb{C}}^{\,\rm S}_x(0)$}{Proof that D = - C'x(0)}}
\label{proof_D}

The prove that $\pmb{D} = -\dotnice{\pmb{C}}^{\rm  S}_x(0)$, we start by considering the discretised form of the Langevin equations \eqref{LE_supp}
\begin{equation}\label{disc_LE}
x^i_{t+dt} - x^i_t = A^i_t dt + \sqrt{2D^{ij}}dW^j_t \, ,
\end{equation}
where $dW^i_t$ is a Wiener process with average and variance given by $\langle dW^i_t \rangle=0$ and $\langle dW^i_t dW^j_t \rangle=\delta^{ij}dt$. By taking the averaged cross product of $i$ and $j$ components of \eqref{disc_LE} one gets
\begin{equation}
    2(C^{\, ij}_{x}(0)-
    \dotnice{C}^{\, (ij)}_{x}(dt)) =  C^{\, ij}_{A}(0)dt^2 + 2 \sqrt{D^{ik}}\sqrt{D^{il}}\delta^{kl}dt \, ,
\end{equation}
where, again, $(ij)$ denotes symmetrized indexes and $C^{\, (ij)}_{O}(t) = (C^{\, ij}_{O}(t)+C^{\, ji}_{O}(t))/2$ is the symmetrized correlation matrix for the observable $\pmb{O}_t$.
Note that correlation functions are homogeneous in time due to the steady-state dynamics of the system. Using further the symmetry of the diffusion matrix $D^{ij}$ and performing a Taylor expansion of $\pmb{C}^{\,(ij)}_{x}(dt)$ for small $dt$ and keeping only terms of order $dt$, it immediately follows that $\pmb{D} = - \dotnice{\pmb{C}}^{\, \rm S}_x(0)$. 

\subsection{Short time behavior for underdamped case} \label{Sec:under_vs_over}

{Let us consider the underdamped Langevin equation in one spatial dimension:
\begin{equation}
\begin{split}
\dot{x}_t &= v_t\,,\\
m \dot{v}_t &= -\gamma v_t + F(x_t) + \sqrt{2D \gamma^2}\, \xi_t\,,
\end{split}
\end{equation}
with $D = k_{\rm B} T/\gamma$ and $\langle \xi_t \xi_s \rangle = \delta(t - s)$. To study the short-time behavior of the position correlation function $C_x(t)$, it is useful to consider the variance of the displacement:
\begin{equation}
\mathrm{Var}(x_t - x_0) = \left\langle \left( \int_0^t v_s\, \dd s \right)^2 \right\rangle = 2(C_x(0) - C_x(t)) \,.
\end{equation}
Using the symmetry $C_v(t-s) = C_v(s-t)$ of the velocity autocorrelation function in the steady state, one obtains:
\begin{equation}
\mathrm{Var}(x_t - x_0) = 2 \int_0^t \int_0^{s'}~ \dd s'  \dd s''\, C_v(s'')\,.
\end{equation}
This implies that:
\begin{equation}
\dotnice{C}_x(0) = 0\,, \qquad \ddotnice{C}_x(0) = -C_v(0)\,.
\end{equation}
Hence, in the underdamped case, $C_x(t)$ has vanishing first derivative at $t=0$, and the leading behavior is quadratic in time, reflecting the expected ballistic short-time motion.}

{In contrast, in the overdamped regime, the velocity is ill-defined and its autocorrelation becomes singular. From the overdamped Langevin equation in \eqref{LE_supp}, one finds that the derivative of $x_t$ satisfies $C_{\dot{x}}(t) \approx 2D\, \delta(t)$ as $t \to 0$. This leads to:
\begin{equation}
    \dotnice{C}_x(0) = -D\,,
\end{equation}
where we have used the identity $\int_0^t \delta(s)\, \dd s = 1/2$, consistent with the Stratonovich convention.}

{This analysis shows that the short-time expansion of the position correlation function distinguishes between overdamped and underdamped dynamics. In particular, the presence or absence of a linear term in $C_x(t)$ near $t=0$ can be used to infer whether inertial effects are present, which is essential for the applicability of our framework and the validity of our main result.
All these arguments extend naturally to multidimensional systems.}

\section{Linear model}\label{lin_mod_disc}

\subsection{Multivariate Ornstein–Uhlenbeck Process}

{As a fundamental example of linear stochastic dynamics, we consider the multivariate Ornstein–Uhlenbeck (OU) process governed by the Langevin equation
\begin{equation} \label{eq:multi_OU}
\dot{\pmb{x}}_t = \pmb{A} \pmb{x}_t + \sqrt{2\pmb{D}}\, \pmb{\xi}_t \, .
\end{equation}
Here, $\pmb{A}$ is the drift matrix. The process is dynamically stable provided all eigenvalues of $\pmb{A}$ have negative real parts, $\Re(\lambda_i) < 0$ for all $i$.}

{The corresponding Fokker–Planck equation is
\begin{equation} \label{multi_OU_FP}
\partial_t p(\pmb{x}, t) = -\nabla \cdot \pmb{j}(\pmb{x}, t) \, ,
\end{equation}
with stationary current and mean local velocity given by
\begin{align} \label{multi_OU_current}
\pmb{j}(\pmb{x}) = \left( \pmb{A} \pmb{x} - \pmb{D} \nabla \right) p(\pmb{x}) \, , \qquad
\pmb{\nu}(\pmb{x}) = \pmb{A} \pmb{x} - \pmb{D} \nabla \log p(\pmb{x}) \, .
\end{align}}

{At steady state, $\partial_t p = 0$, so $\nabla \cdot \pmb{j} = 0$. The distribution is Gaussian with zero mean and covariance matrix $\pmb{C}_x(0)$, which satisfies the Lyapunov equation
\begin{equation} \label{eq:lyapunov}
\pmb{A} \pmb{C}_x(0) + \pmb{C}_x(0) \pmb{A}^\top = -2\pmb{D} \, .
\end{equation}}

{At equilibrium, the condition $\pmb{\nu} = 0$ implies $\pmb{A} = -\pmb{D} \pmb{C}_x^{-1}(0)$, which automatically satisfies the Lyapunov equation. In the steady state, the pdf satisfies 
$p(\pmb{x}) = \mathcal{Z}^{-1} \exp\left(-\frac{1}{2} \pmb{x}^{\rm \, T} \pmb{C}_x^{-1}(0) \pmb{x} \right)$, 
and the mean local velocity becomes 
$\pmb{\nu}(\pmb{x}) = \left( \pmb{A} + \pmb{D} \pmb{C}_x^{-1}(0) \right) \pmb{x}$.}

{It is useful to decompose the drift matrix as $\pmb{A} = \pmb{\mu}(-\pmb{K} + \pmb{\psi})$, where $\pmb{K}$ is a symmetric, positive-definite matrix derived from a quadratic conservative potential, and $\pmb{\psi}$ is an antisymmetric matrix representing non-conservative forces. This decomposition is always possible, since any square matrix can be uniquely decomposed into symmetric and antisymmetric parts. In practice, non-conservative forces may not be purely rotational but simply non-reciprocal, in which case the decomposition can be understood as an effective representation in terms of $\pmb{K}$ and $\pmb{\psi}$.}

{In terms of this decomposition, the inflow becomes:
\begin{equation}
\inflow = -\int d\pmb{x}_t \, p(\pmb{x}_t) \, \partial_i A^i_t = -\int d\pmb{x}_t \, p(\pmb{x}_t) \, A^{ij} \delta^{ij} = -\mathrm{Tr}(\pmb{A}) = \mathrm{Tr}(\pmb{\mu} \pmb{K}) \, ,
\end{equation}
which depends only on the symmetric part of the drift matrix,
since $\mathrm{Tr}(\pmb{\mu} \pmb{\psi}) = 0$. Indeed, for any antisymmetric matrix $\pmb{\psi}$ and symmetric matrix $\pmb{\mu}$, the trace of their product vanishes. This shows that the inflow is independent of the rotational part of the drift and also insensitive to temperature differences, i.e. the sources of non-equilibrium. If $\pmb{D}$ is diagonal, this result extends to the case where $\pmb{\psi}$ is simply non-symmetric with zero diagonal, since diagonal terms can be absorbed into $\pmb{K}$ without loss of generality.}

{Using $\nabla \log p(\pmb{x}) = -\pmb{C}_x^{-1}(0) \pmb{x}$ and the expression for the local velocity in Eq.~\eqref{multi_OU_current}, one can also show that the entropy production rate is equal to
\begin{equation}
\sigma = \langle \pmb{\nu}^\top \pmb{D}^{-1} \pmb{\nu} \rangle = \mathrm{Tr}\left( \pmb{D}^{-1} \pmb{A} \pmb{C}_x(0) \pmb{A}^\top \right) - \inflow \, .
\end{equation}}

\subsection{2D model}\label{2D_lin_mod_sol}

In this section we analytically solve the linear 2D system presented in the main text with dynamical equations given by
\begin{equation} \label{SDE_linear_supp}
    \begin{split}
        \dot{x}^1_t &= A_{11}\,x^1_t+A_{12}\,x^2_t+\sqrt{2T}\,\xi^1_t \, ,\\[3pt]
        \dot{x}^2_t &= A_{22}\,x^2_t+\alpha_A A_{12}\,x^1_t+\sqrt{2 \alpha_TT}\,\xi^2_t \, ,
    \end{split}
\end{equation}

\subsubsection{Equal time correlation matrix}

The first step consists of calculating the components of the equal-time correlation matrix $\pmb{C}_{x}(0)$ which can be done by considering the discretized form of Eq. \eqref{SDE_linear_supp}, which reads
\begin{equation} \label{SDE_linear_supp_disc}
    \begin{split}
        x^1_{t+dt} &= x^1_{t}+ \left(A_{11}\,x^1_t+A_{12}\,x^2_t\right)dt+\sqrt{2T}\,dW^{1}_t \, ,\\[3pt]
        x^2_{t+dt} &= x^2_{t}+\left(A_{22}\,x^2_t+\alpha_A A_{12}\,x^1_t\right)dt+\sqrt{2 \alpha_TT}\,dW^{2}_t \, ,
    \end{split}
\end{equation}
where $dW^i_t$ represents a Wiener process characterized by a mean of $\langle dW^i_t \rangle = 0$ and a variance of $\langle dW^i_t dW^j_t \rangle = \delta^{ij} dt$. By squaring and taking the cross product of the equations in \eqref{SDE_linear_supp_disc}, followed by averaging and retaining only terms of order $\mathcal{O}(dt)$, we obtain:
\begin{equation}\label{lin_system_step_1}
    \begin{split}
        \langle (x^1_{t+dt})^2\rangle =& \langle (x^1_{t})^2 \rangle  + 2(A_{11}\langle (x^1_t)^2\rangle+A_{12}\langle x^1_t x^2_t \rangle )dt+2Tdt\,, \\
         \langle (x^2_{t+dt})^2\rangle =& \langle (x^2_{t})^2 \rangle  + 2(A_{22}\langle (x^2_t)^2\rangle+\alpha_{A}A_{12}\langle x^1_t x^2_t  \rangle )dt+2\alpha_T T dt \,,\\
         \langle x^1_{t+dt}x^2_{t+dt}\rangle  =& \langle x^1_{t}x^2_{t} \rangle  + (A_{12}\langle (x^2_t)^2 \rangle+\alpha_{A}A_{12}\langle (x^1_t)^2 \rangle + (A_{11}+A_{22})\langle x^1_t x^2_t\rangle )dt\,,
    \end{split}
\end{equation}
where we used that $\langle O_t \, dW^i_t \rangle =0 $, $O_t=x^1_t,x^2_t$. Note these three equations univocally determine the components of $\pmb{C}_{x}(0)$ which is symmetric in a NESS. In particular, 
\begin{equation}\label{lin_system_step_2}
\begin{split}
&A_{11} C^{\,11}_x(0)+A_{12} C^{\,12}_x(0) = T\,,\\[3pt]
&A_{22} C^{\,22}_x(0)+\alpha_{A}A_{12} C^{\,12}_x(0) = \alpha_T T\,,\\[3pt]
A_{12}(C^{\,11}_x(0)&+\alpha_{A}C^{\,22}_x(0))+(A_{11}+A_{22})C^{12}_x(0) =0 \,,
    \end{split}
\end{equation}
whose solution is:
\begin{equation}
\begin{split}\label{eq:asym_2D_eq_time_corr}
        C^{\,11}_{x}(0) &= \frac{T \left(\alpha_{A}A^2_{12} - A_{22} (A_{11} + A_{22}) - \alpha TA_{12}^2  \right)}{(A_{11} + A_{22}) \left( A_{11} A_{22}- \alpha_{A}A^2_{12}  \right)} \, ,\\[5pt]
        C_{x}^{\,22}(0) &= \frac{T \left( \alpha T( \alpha_{A}A^2_{12} - A_{11} (A_{11} + A_{22}))  -\alpha_{A}^2A_{12}^2\right)}{(A_{11} + A_{22}) \left( A_{11} A_{22} - \alpha_{A} A_{12}^2 \right)}\, ,\\[5pt]
        C_{x}^{\,12}(0) &= C_{x}^{\,21}(0) =\frac{T \left(\alpha_{A} A_{22} A_{12}  + \alpha T A_{11} A_{12}  \right)}{(A_{11} + A_{22}) \left( A_{11} A_{22} - \alpha_{A}A_{12}^2 \right)}\, ,\\[4pt]
    \end{split}
\end{equation}
namely the components of the equal time correlation matrix $\pmb{C}_x(0)$. 

\subsubsection{Time dependent correlation matrix}

To compute the time-dependent correlation matrix $\pmb{C}_x(t)$, we multiply each equation in \eqref{SDE_linear_supp}, evaluated at a time $t=0$, by $x^1_t$ and $x^2_t$ evaluated at $t>0$, and then take the ensemble average, resulting in:
\begin{equation}\label{eq:asym_2D_corr_time}
    \begin{split}
        & \partial_t C^{\,11}_{x}(t) = A_{11}C^{\,11}_{x}(t)+A_{12}C^{\,21}_{x}(t)\,, \\
        & \partial_t C^{\,12}_{x}(t) = A_{11}C^{\,12}_{x}(t)+A_{12}C^{\,22}_{x}(t)\,, \\
        & \partial_t C^{\,21}_{x}(t) = A_{22}C^{\,21}_{x}(t)+\alpha_{A}A_{12}C^{\,11}_{x}(t)\,,\\
        & \partial_t C^{\,22}_{x}(t) = A_{22} C^{\,22}_{x}(t)+\alpha_{A}A_{12} C^{\,12}_{x}(t) \, ,
    \end{split}
\end{equation}
where we used that $\langle O_t \chi_0 \rangle=0$ for $t>0$. To solve this linear system of differential equations, we resort to standard procedure based on Laplace transform, which leads to 
\begin{equation}
\begin{split}\label{eq:asym_2D_corr_laplace_sol}
        & \widehat{C}^{\, 11}_{x}(s) = \frac{C^{\, 11}_x(0) (s-A_{22} )+A_{12} C^{\,12}_{x}(0) }{(s-A_{11} ) (s-A_{22})- \alpha_{A}A^2_{12}}\,,\hspace{1.7cm} 
        \widehat{C}^{\,12}_{x}(s) = \frac{C^{\,12}_{x}(0) (s-A_{22})+A_{12} C^{\,22}_{x}(0) }{(s-A_{11}) (s-A_{22})- \alpha_{A}A^2_{12}}\,,\\[5pt]
        & \widehat{C}^{\,21}_{x}(s) = \frac{C^{\,12}_{x}(0) (s-A_{11})+\alpha_{A}A_{12} C^{\,11}_{x}(0) }{(s-A_{11}) (s-A_{22})- \alpha_{A}A^2_{12} }\,,\hspace{1.3cm}
         \widehat{C}^{\,22}_{x}(s) = \frac{C^{\,22}_{x}(0) (s-A_{11})+\alpha_{A}A_{12} C^{\,12}_{x}(0) }{(s-A_{11}) (s-A_{22} )- \alpha_{A}A^2_{12}}\, .\\[4pt]
    \end{split}
\end{equation}
where $\widehat{\pmb{C}}_{x}(s)$ is the Laplace transform of $\pmb{C}_{x}(s)$.
In the time domain, the corresponding expressions are:
\begin{eqnarray}\label{time_corr_sol}
    C^{\,11}_{x}(t) &=& e^{(A_{11}+A_{22})t/2} \left( C^{\,11}_{x}(0) \cosh\left(t\Delta \right) + \frac{(A_{11} - A_{22})C^{\,11}_{x}(0) + 2 A_{12} C^{\,12}_{x}(0)}{2 \Delta} \sinh\left(t\Delta \right) \right) \,,\nonumber \\
    C^{\,12}_{x}(t) &=& e^{(A_{11}+A_{22})t/2} \left( C^{\,12}_{x}(0) \cosh\left(t\Delta \right) + \frac{(A_{11} - A_{22})C^{\,12}_{x}(0) + 2 A_{12} C^{\,22}_{x}(0)}{2 \Delta} \sinh\left(t\Delta \right) \right) \,,\\
    C^{\,21}_{x}(t) &=& e^{(A_{11}+A_{22})t/2} \left( C^{\,21}_{x}(0) \cosh\left(t\Delta \right) + \frac{(A_{22} - A_{11})C^{\,21}_{x}(0) + 2 \alpha_{A}A_{12} C^{\,11}_{x}(0)}{2 \Delta} \sinh\left(t\Delta \right) \right) \,,\nonumber \\
    C^{\,22}_{x}(t) &=& e^{(A_{11}+A_{22})t/2} \left( C^{\,22}_{x}(0) \cosh\left(t\Delta \right) + \frac{(A_{22} - A_{11})C^{\,22}_{x}(0) + 2 \alpha_{A}A_{12} C^{\,21}_{x}(0)}{2 \Delta} \sinh\left(t\Delta \right) \right) \,,\nonumber
\end{eqnarray}
where $\Delta =  \sqrt{ \alpha_{A}A^2_{12} + ((A_{11} - A_{22})/2)^2}$. By combining \eqref{eq:asym_2D_eq_time_corr} with \eqref{time_corr_sol}, one can readily evaluate first and second time derivative of the correlation matrix at time $t=0$. In particular, for first derivatives one gets
\begin{equation}\label{lin_first_der}
    \dotnice{C}^{\,11}_{x}(0) = -T \, , \quad \dotnice{C}^{\,12}_{x}(0) =  \frac{T(\alpha_{A}A_{12}  - \alpha_T A_{12}) }{A_{11} + A_{22}} \, , \quad \dotnice{C}^{\,21}_{x}(0) =  \frac{T( \alpha_T A_{12} - \alpha_{A}A_{12}) }{A_{11} + A_{22}} \, , \quad \dotnice{C}^{\,22}_{x}(0) = -\alpha_{T} T \, ,
\end{equation}
for which it holds that $\pmb{D} = -\dotnice{\pmb{C}}^{\, \rm S}_x(0)$, as expected. Finally, for second derivatives, one gets:
\begin{equation}
\begin{split}\label{eq:lin_sec_der}
        & \ddotnice{C}^{\, 11}_{x}(s) = -T A_{11} + \frac{T(\alpha_T-\alpha_A ) A_{12}^2 }{A_{11} + A_{22}}
 \, ,
\hspace{2.3cm} 
        \ddotnice{C}^{\,12}_{x}(s) = \frac{ T A_{12} \left(\alpha_A A_{11}  - \alpha_T(2 A_{11} + A_{22}) \right)}{A_{11} + A_{22}}
 \, ,
\\[5pt]
        & \ddotnice{C}^{\,21}_{x}(s) = \frac{ T A_{12}\left(\alpha_T A_{22} -\alpha_A(A_{11} + 2 A_{22}) \right)}{A_{11} + A_{22}}
 \, ,
\hspace{1.7cm}
         \ddotnice{C}^{\,22}_{x}(s) = - T \alpha_T A_{22} + \frac{ T \alpha_A (\alpha_A - \alpha_T)A_{12}^2}{A_{11} + A_{22}}  
\, .\\[4pt]
    \end{split}
\end{equation}
Using all these results, one can straightforwardly calculate the traffic
\begin{equation}
     4\traffic = -{\rm Tr}\!\left[\pmb{D}^{-1} \,\ddotnice{\pmb{C}}^{\, \rm S}_x(0)\right] = -\frac{\ddotnice{C}^{\, 11}_{x}(s)}{T}-\frac{\ddotnice{C}^{\, 22}_{x}(s)}{\alpha_T T} = -(A_{11}+A_{22}) -\frac{A_{12}^2(\alpha_A-\alpha_T)^2}{\alpha_T(A_{11}+A_{22})}
\end{equation}

\subsubsection{Effective forces and inflow rate}

Effective forces in a NESS can be easily estimated for this linear system as the probability distribution is Gaussian,
\begin{equation}
    p(\pmb{x}) = \frac{1}{\sqrt{2\pi |\pmb{C}_{x}(0)|}} \exp\left(-\frac{1}{2} \pmb{x}^{\rm T} \pmb{C}^{\rm-1}_{x}(0) \,\pmb{x}\right)\, ,
\end{equation}
where $|\cdot|$ stands for the determinant. Hence, the effective forces $-\nabla \phi(\pmb{x}) = \nabla\log p(\pmb{x}_t) $ become
\begin{equation}
\begin{split}
   \partial^1_t \phi(\pmb{x}) =&\frac{(A_{11} + A_{22}) \left(\alpha_A  (\alpha_A - \alpha_T) A_{12}^2 x_t^1 +\alpha_T A_{11} (A_{11} + A_{22}) x_t^1  + A_{12}  \left(\alpha_A A_{22}  + \alpha_T A_{11}  \right) x_t^2 \right)}{T \left(A_{12}^2 (\alpha_A - \alpha_T)^2 + \alpha_T (A_{11} + A_{22})^2 \right)}
\, ,
\\[5pt]
   \partial^2_t \phi(\pmb{x}) =& \frac{(A_{11} + A_{22}) \left(A_{22}^2 x_t^2 + \alpha_A A_{12} A_{22} x_t^1  + (\alpha_T -\alpha_A )A_{12}^2   x_t^2 + A_{11} \left(A_{22} x_t^2 + \alpha_T A_{12} x_t^1 \right)\right)}{T \left((\alpha_A - \alpha_T)^2 A_{12}^2  + \alpha_T (A_{11} + A_{22})^2 \right)}
\, .
\end{split}
\end{equation}
With some algebraic manipulation, this enables the calculation of the correlation matrix for the effective forces, $\pmb{C}_{\nabla \phi}(0)$, which is essential for determining the inflow rate, $\inflow = {\rm Tr}\!\left[\pmb{D} \pmb{C}_{\nabla \phi}(0)\right]$.
The components of the symmetric matrix $\pmb{C}_{\nabla \phi}(0)$ are given by:
\begin{equation}
\begin{split}
     C^{\, 11}_{\nabla \phi}(0) = & -\frac{(A_{11} + A_{22}) \left(\alpha_T A_{11} (A_{11} + A_{22}) +\alpha_A (\alpha_A - \alpha_T) A_{12}^2  \right)}{T \left((\alpha_A - \alpha_T)^2 A_{12}^2  + \alpha_T(A_{11} + A_{22})^2  \right)}\, ,\\[5pt]
     C^{12}_{\nabla \phi}(0) = &\,  C^{21}_{\nabla \phi}(0) = -\frac{A_{12} (A_{11} + A_{22}) \left(\alpha_T A_{11} + \alpha_AA_{22}  \right)}{T \left((\alpha_A - \alpha_T)^2 A_{12}^2  + \alpha_T (A_{11} + A_{22})^2  \right)}\, ,
     \\[5pt]
     C^{\, 22}_{\nabla \phi}(0) = & -\frac{(A_{11} + A_{22}) \left(A_{22} (A_{11} + A_{22}) + (\alpha_T-\alpha_A ) A_{12}^2 \right)}{T \left((\alpha_A - \alpha_T)^2 A_{12}^2  + \alpha_T (A_{11} + A_{22})^2  \right)} \, .
 \\[5pt]  
\end{split}
\end{equation}
Because the diffusion matrix $\pmb{D}$ is diagonal, with $D^{11} = T$ and $D^{11}=\alpha_{T} T$, the inflow rate $\inflow$ reduces to:
\begin{equation}
    \inflow = {\rm Tr}\!\left[\pmb{D} \pmb{C}_{\nabla \phi}(0)\right] = T C^{\, 11}_{\nabla \phi}(0) +\alpha_T T C^{\, 22}_{\nabla \phi}(0)= -(A_{11}+A_{22}) \, ,
\end{equation}
    as anticipated in the main text.

\subsubsection{Entropy production}

The entropy production rate $\sigma$ associated to the system can be calculated using the standard formula in Eq.~\eqref{sekimoto} presented in the main text. By identifying the forces acting on the the degrees of freedom $\{x^1_t,x^2_t\}$
\begin{align}
    F^{1}_t = A_{11} x^1_t + A_{12}x^2_t && F^{2}_t = A_{22} x^2_t + \alpha_{A} A_{12} x^1_t
\end{align}
Eq.~\eqref{sekimoto} can be readily be evaluated as 
\begin{equation}\label{supp_ent_lin_step1}
    \sigma = \frac{1}{k_{\rm B}T}\langle F^1_t\circ\dot{x}^1_t \rangle + \frac{1}{\alpha_T k_{\rm B} T}\langle F_t^2\circ\dot{x}^2_t \rangle = \frac{A_{12}}{k_{\rm B}T} \left( \langle x^2_t\circ\dot{x}^1_t\rangle + \frac{\alpha_A}{\alpha_T} \langle x^1_t\circ\dot{x}^2_t\rangle  \right)\, , 
\end{equation}
because $\langle x^i_t\circ\dot{x}^i_t\rangle = \partial_t  \langle (x^i_t)^2\rangle =0 $ in a NESS. By further plugging in the dynamical equations form \eqref{SDE_linear_supp} into \eqref{supp_ent_lin_step1} one finally gets gets
\begin{equation}
    \sigma = \frac{ A^2_{12}}{k_{\rm B} T} \left( C^{\,22}_{x}(0) + \frac{\alpha_{A}^2}{\alpha_{T}} C^{\,11}_{x}(0)\right) + \frac{A_{12}}{k_{\rm B} T} \left( A_{11} + \frac{\alpha_{A}}{\alpha_{T}} A_{22}\right) C^{\,12}_{x}(0) = -\frac{A_{12}^2(\alpha_A-\alpha_T)^2}{\alpha_T(A_{11}+A_{22})}\, ,
    \label{eqS:totalep}
\end{equation}
where we used the results in \eqref{eq:asym_2D_eq_time_corr} and that $\langle x^i_t \circ \xi^j_t \rangle = \sqrt{D^{ij}/2} $ only has diagonal terms, as shown in Section \ref{proof_strat_corr}.

\section{\texorpdfstring{Proof that $\langle x^i_t \circ \xi^j_t \rangle = \sqrt{D^{ij}/2}$}{Proof that strat corr}}
\label{proof_strat_corr}

In this section, we aim to compute the correlation $ \langle x^i_t \circ \xi^j_t \rangle $, where the symbol $ \circ $ denotes the Stratonovich product, for a system of Langevin equations as in Eq.~\eqref{LE_supp}:
\begin{equation}
\dot{x}^i_t = A^i_t + \sqrt{2D^{ik}} \, \xi^k_t \, ,
\end{equation}
with $\langle \xi^i_t \rangle = 0$ and $\langle \xi^i_t \, \xi^j_s \rangle = \delta^{ij} \delta(t-s)$. In the Stratonovich interpretation, the stochastic integral evaluates $ x^i_t $ at the midpoint of the integration interval, introducing a correlation between the variable $ x^i_t $ and the noise $ \xi^j_t $.
To compute $ \langle x^i_t \circ \xi^j_t \rangle $, we use the formal solution of the Langevin equation:
\begin{equation}
x^i_t = x^i_0 + \int_0^t \dd t^{\prime} A^i_{t^{\prime}} + \sqrt{2D^{ik}} \int_0^t \dd t^{\prime}\,\xi^k_{t^{\prime}} \, .
\end{equation}
At short times, the dominant contribution to $ x^i_t $ arises from the noise term as $\xi^{i}_t \sim dt^{-1/2}$, while the deterministic term contributes only to longer timescales. Thus, for small $ dt $, the dynamics of $ x^i_t $ can be approximated as:
\begin{equation}
x^i_t \approx x^i_0 + \sqrt{2D^{ik}} \int_0^t \dd t^{\prime} \xi^k_{t^{\prime}} \, .
\end{equation}
Using this approximate form of $ x^i_t $, and by multiplying both sides by $\circ \,\xi^i_t$, one gets:
\begin{equation}\label{eq:strat_corr_step1}
\langle x^i_t \circ \ \xi^j_t \rangle =  \sqrt{2D^{ik}}  \int_0^t \, \dd t^{\prime}\langle \xi^k_{t^{\prime}} \circ \xi^j_t \rangle = \sqrt{2D^{ik}} \,\delta^{kj} \!\int_0^t \, \dd t^{\prime} \delta(t-t')\, ,
\end{equation}
where we used $\langle x^i_0 \circ \ \xi^j_t \rangle=0$. 
Within the Stratonovich convention, integrating the Dirac delta function centered at one of the integration limits gives
\begin{equation}
   \int_0^t \, \dd t^{\prime} \delta(t-t') f(t') = f(t)/2 \,
\end{equation}
which finally implies that \eqref{eq:strat_corr_step1} becomes
\begin{equation}
    \langle x^i_t \circ \ \xi^j_t \rangle = \sqrt{D^{ij}/2}  \, .
\end{equation}

\section{Non existence of equilibrium solutions for 2D linear model}\label{non_exist_sol}

\begin{figure*}[t!]
 	\centering
\includegraphics[width=0.45\textwidth]{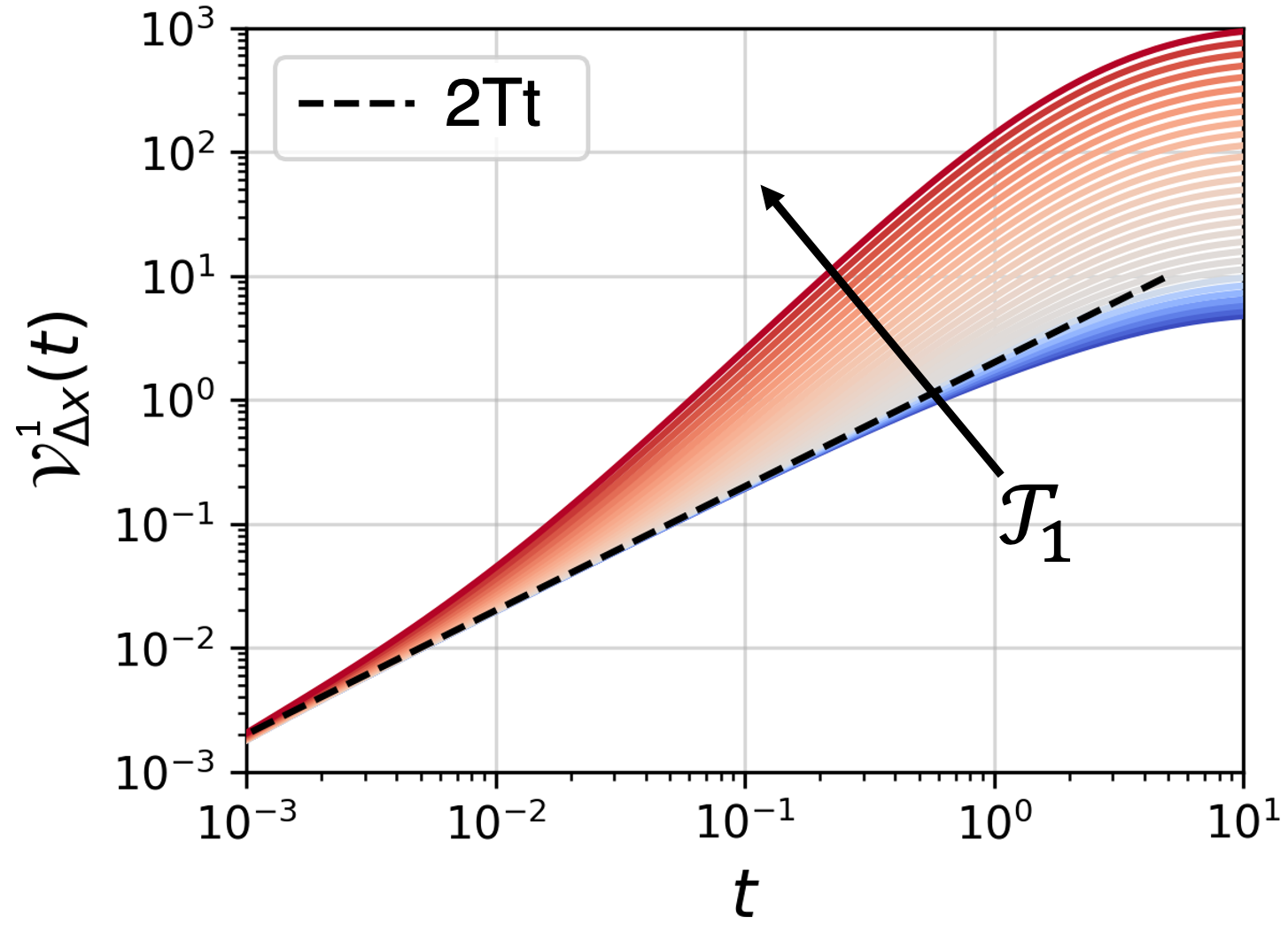}
	\caption{Plot of ${\cal V}^{\, 1}_{\Delta x}(t) = {\rm Var}(x^{1}_t-x^1_0) =  2 (C_{x}^{11}(0)-C_{x}^{11}(t))$ with $A_{11} = -1$, $A_{22}=-1$, $A_{12}=1$, $\alpha_A=0.5$, $T=1$ and $\alpha_T \in [1,500]$. Note the deviation from normal diffusion as  $\traffic_1$ (and hence $\sigma$) increases, validating enhanced diffusion as a signature of nonequilibrium in overdamped systems.}
	\label{Fig_supp_sec_der}
\end{figure*}

The problem of detecting the nonequilibrium nature of a one-dimensional Gaussian stochastic process can be effectively addressed within the framework of the 2D linear model \eqref{SDE_linear_supp} discussed in this paper. Following the approach in \cite{lucente2022inference}, we consider the scenario where only one of the two degrees of freedom (DOFs) in \eqref{SDE_linear_supp} is observed, for example $x^1_t$. This implies that only the autocorrelation function $C^{\, 11}_{x}(t)$ can be inferred from the data. 
Since the system is Gaussian, the propagator of the associated Fokker-Planck equation is also Gaussian, meaning that the system is fully characterized by the two-point correlation function. Assuming $C^{\, 11}_{x}(t)$ is computed and the data is modeled by \eqref{SDE_linear_supp}, one could attempt to fit the model parameters and infer $\sigma$ directly, hence determining whether or not the process is irreversible or not. However, even if the system is out of equilibrium, there exist certain parameter regimes where inferring nonequilibrium behavior becomes impossible. Specifically, we will show that in regimes where the correlation function $C^{\, 11}_{x}(t)$ exhibits positive concavity (i.e., $\traffic_1 \leq 0$), there always exists a set of parameters for which the system appears to be in equilibrium, reproducing the exact measured correlation function. This degeneracy renders the inference of nonequilibrium behavior ambiguous. We further demonstrate that when $\traffic_1 > 0$, in agreement with our result \eqref{sigma_part}, detection of the nonequilibrium nature of a 1D Gaussian stochastic process becomes feasible. This arises because in such cases, no parameter set exists that generates a system at equilibrium while remaining compatible with the observed correlation function. To establish this, we express $C^{\, 11}_{x}(t)$ in Fourier space, leveraging the relation between Laplace and Fourier transforms:
\begin{equation}
    \widetilde{C}^{\, 11}_{x}(\omega) = \widehat{C}^{\, 11}_{x}(i\omega) + \widehat{C}^{\, 11}_{x}(-i\omega) = \frac{c_0 + \omega^2 c_1}{|(i\omega)^2 + i\omega {\rm Tr} + {\cal D}|^2} \, ,
\end{equation}
where we utilize Eq.~\eqref{eq:asym_2D_corr_laplace_sol}. The coefficients, which can be experimentally estimated from the data, are given by:
\begin{equation}\label{system_pars_2D}
\begin{aligned}
    c_0 &= 2T(\alpha_{T}A_{12}^2 + A_{22}^2 ), \\
    c_1 &= 2T, \\
    {\rm Tr} &= A_{11} + A_{22}, \\
    {\cal D} &= A_{11}A_{22} - \alpha_{A}A_{12}^2 \, ,
\end{aligned}
\end{equation}
with ${\rm Tr} < 0$, and ${\cal D} > 0$ to ensure the existence of a nonequilibrium steady state (NESS). This implies that any combination of $A_{11}$, $A_{22}$, $A_{12}$, $\alpha_{A}$, $T$, and $\alpha_{T}$ producing the observed values in \eqref{system_pars_2D} is a potential candidate for the true set of system parameters. We now investigate the conditions under which a parameter set yielding an equilibrium system can solve the equations in \eqref{system_pars_2D}. Assuming the two DOFs in \eqref{SDE_linear_supp} are coupled (i.e., $A_{12} \neq 0$), the system can be at equilibrium if and only if $\alpha_A = \alpha_T \equiv \alpha >0$, as $\alpha_T$ must be positive. Under this condition, \eqref{system_pars_2D} simplifies to:
\begin{equation}\label{system_pars_2D_eq}
\begin{aligned}
    c_0 &= 2T(\alpha A_{12}^2 + A_{22}^2), \\
    c_1 &= 2T, \\
    {\rm Tr} &= A_{11} + A_{22}, \\
    {\cal D} &= A_{11}A_{22} - \alpha A_{12}^2 \, .
\end{aligned}
\end{equation}

Through algebraic manipulation, we find that this leads to the following condition for $A_{11}$:
\begin{equation}\label{A11_condition}
    A_{11} = {\rm Tr} - \frac{1}{{\rm Tr}} \left({\cal D} + \frac{c_0}{c_1}\right) \, ,
\end{equation}
which must satisfy $A_{11} < 0$ to guarantee the stability of the equilibrium solution. Indeed, if $A_{11} \ge 0$, then $A_{22}$ must be negative to ensure ${\rm Tr} = A_{11}+A_{22}<0$. But this would mean that $\mathcal{D}<0$, as $\alpha \ge 0$, ruling out a stable equilibrium. As a consequence, if for example the measured $c_0$ satisfies:
\begin{equation}
    c_0 \geq c_1 \left( {\rm Tr}^2 - {\cal D} \right) \, ,
\end{equation}
which could occur with a sufficiently large $\alpha_T$ (as it does not affect $c_1$, ${\cal D}$, or ${\rm Tr}$), this would result in $A_{11} > 0$. This disrupts the existence of a stationary equilibrium, meaning that imposing equilibrium conditions on the system in \eqref{system_pars_2D_eq} leads to a contradiction and therefore the observed correlation function is incompatible with equilibrium dynamics.
More generally, this incompatibility arises whenever the right-hand side of \eqref{A11_condition} is positive, which corresponds to:
\begin{equation}
    0 \leq {\rm Tr} - \frac{1}{{\rm Tr}} \left({\cal D} + \frac{c_0}{c_1}\right) =  A_{11} - \frac{(\alpha_T-\alpha_A ) A_{12}^2 }{A_{11} + A_{22}} = 4\traffic_1 \, ,
\end{equation}
as can be shown using the expressions for the first and second derivatives of $C^{\, 11}_{x}(t)$ in \eqref{lin_first_der} and \eqref{eq:lin_sec_der}, along with the definition of traffic components, namely $4 \traffic_1 = -\ddotnice{C}^{\, 11}_x(0) / \dotnice{C}^{\, 11}_x(0)$. In other words, because at equilibrium \(4\,\traffic_1 = A_{11}\), a measured \(\traffic_1 < 0\) can still yield \(A_{11} \le 0\), thus keeping the system compatible with equilibrium. Conversely, if \(\traffic_1 > 0\), the condition \(A_{11} < 0\) cannot be satisfied, indicating that the system must be out of equilibrium.

To conclude this section, we visually illustrate in Fig.~\ref{Fig_supp_sec_der} how the behavior of the displacement variance, ${\cal V}^{\, 1}_{\Delta x}(t) = 2(C^{\, 11}_{x}(0) - C^{\, 11}_{x}(t))$, evolves as $\traffic_1$ increases. This term corresponds to the first component on the right-hand side of the VSR \eqref{eq:VSR1}. As $\traffic_1$ increases, we observe an enhanced diffusion compared to the baseline provided by normal diffusion (represented by the solid black line).


\section{Detectability of non-equilibrium in linear models}

{A key feature of the method introduced in the main text is its ability to estimate entropy production using only a subset of observable degrees of freedom. However, the quality and informativeness of this estimate can vary significantly depending on how energy dissipation is distributed across the system. In particular, the detectability of entropy production from traffic depends on the position of the observed variables relative to the sources of nonequilibrium driving.}

{To explore this issue systematically, we first consider again the a minimal 2D linear model presented in the main text. This allows us to study how the traffic components and their relation to the total entropy production depend on parameters such as temperature imbalance and coupling asymmetry. The insights gained here help establish when partial observations suffice to infer dissipation and when model-based approaches may become necessary. We then extend this analysis to the more complex three-dimensional model used in \cite{VSR}, where only one component is usually observable and model-free inference becomes more challenging.}

\subsection{2D model}

{We start by considering the 2D linear model:
\begin{equation} \label{SDE_linear_resp}
\begin{split}
    \dot{x}^1_t &= A_{11}\,x^1_t + A_{12}\,x^2_t + \sqrt{2T_1}\,\xi^1_t \,,\\[3pt]
    \dot{x}^2_t &= A_{22}\,x^2_t + A_{21}\,x^1_t + \sqrt{2T_2}\,\xi^2_t \,,
\end{split}
\end{equation}
which offers important insights into how traffic reflects the underlying entropy production rate $\sigma$. As shown in Fig.~3a of the main text (where $T_1 = T$ and $T_2 = \alpha_T T$), increasing the temperature imbalance in favor of the hidden degree of freedom $x^2_t$ (i.e., increasing $\alpha_T$) leads to higher traffic in the observed component $x^1_t$, $\mathcal{T}_1$. This suggests that when hidden degrees of freedom inject heat into colder and observable components, the resulting dissipation becomes more easily detectable from the latter.}

\begin{figure*}[t!]
 	\centering
\includegraphics[width=1\textwidth]{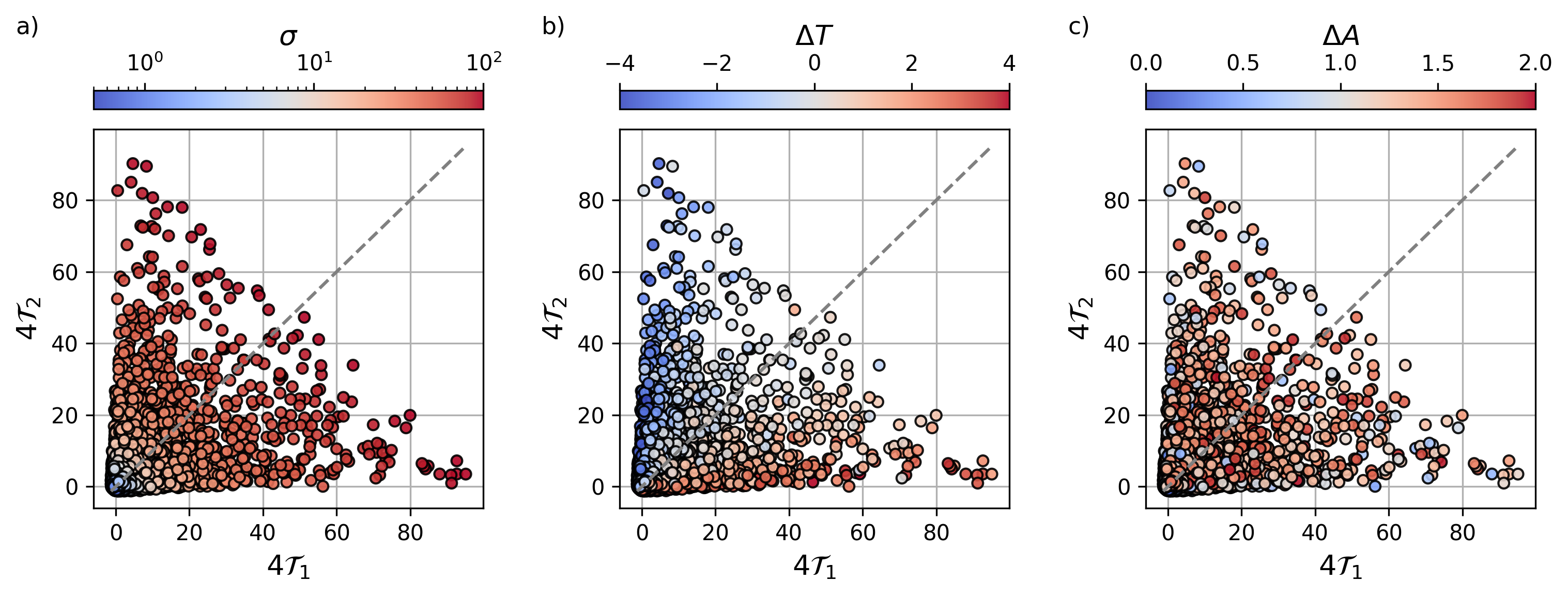}
	\caption{10000 realizations of the 2D model in Eq.~\eqref{SDE_linear_resp} with randomly sampled parameters in the following intervals: $A_{11} \in (-1, 1)$, $A_{22} \in (-1, 1)$, $A_{12} \in (-1, 1)$, $A_{21} \in (-1, 1)$, $T_1 \in (0.1, 2)$, $T_{2} \in (0.1, 2)$.}  
	\label{Fig_resp_4}
\end{figure*}

\begin{figure*}[t!]
 	\centering
\includegraphics[width=0.7\textwidth]{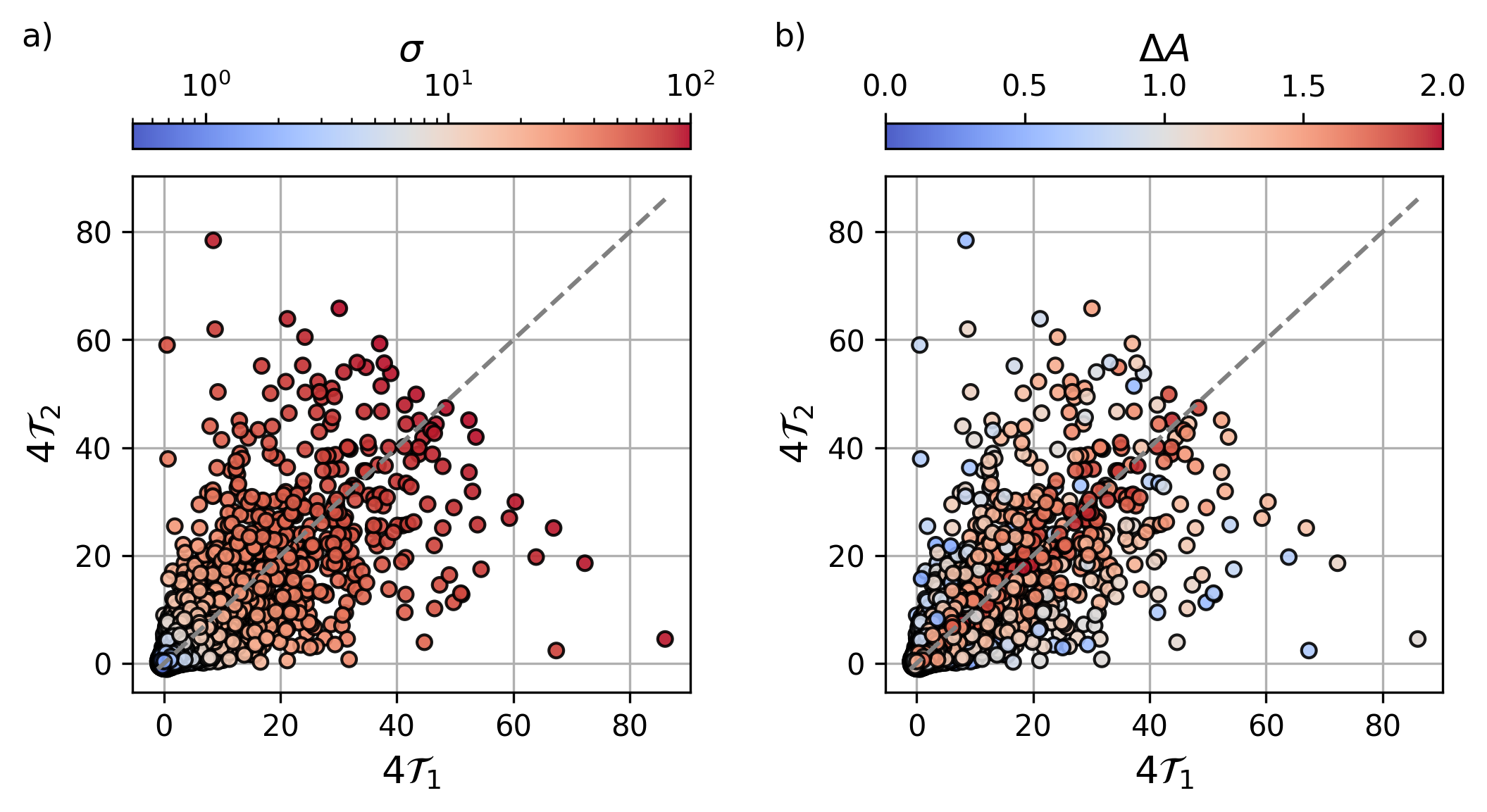}
	\caption{10000 realizations of the 2D model as in Fig.~\ref{Fig_resp_4} but fixing $T_1=T_2 \in (0.1, 2)$.}  
	\label{Fig_resp_5}
\end{figure*}

{To support this point, we computed $\sigma$, $\mathcal{T}_1$, and $\mathcal{T}_2$ across $10^4$ randomly sampled parameter sets of the system above, constrained by the stability conditions $A_{11}A_{22} - A_{12}A_{21} > 0$ and $A_{11} + A_{22} \le 0$. To analyze how detectability varies, we defined two quantities: $dt = (T_2 - T_1)/\sqrt{T_1 T_2}$ for the temperature imbalance, and $\Delta A = |A_{12} - A_{21}|$ for the asymmetry in mechanical coupling. Results are shown in Fig.~\ref{Fig_resp_4}. In panel a), we plot $4\mathcal{T}_1$ on the x-axis and $4\mathcal{T}_2$ on the y-axis, with the color bar indicating the total entropy production $\sigma$. In panel b), the same axes are used, but the color scale represents the temperature imbalance $dt$, while in panel c), the color scale corresponds to the coupling asymmetry $\Delta A$. This visualization allows us to assess how the detectability of entropy production shifts depending on where dissipation is being injected and how coupling is structured. Panel b) of Fig.~\ref{Fig_resp_4} shows that detectability of $\sigma$ from $x^1$ improves with increasing $dt$, i.e., when $T_2 > T_1$, hence making $\sigma$ more detectable when observing $x^1$ rather than $x^2$. Conversely, when $dt$ is negative, detectability shifts to $x^2$. This confirms that entropy production becomes more visible from a degree of freedom that receives heat from others. As for mechanical asymmetry $\Delta A$, its effect on the distribution of $\sigma$ between $\mathcal{T}_1$ and $\mathcal{T}_2$ is minor when temperature imbalance is present (see Fig.~\ref{Fig_resp_4}c), but becomes more pronounced when $dt = 0$. In that case, entropy production is shared more evenly between the two DOFs (see Fig.~\ref{Fig_resp_5}b). This behavior is also reflected in the hair-cell model: although there is a small effective temperature difference ($T^{\rm eff} \approx 1.5T$, see Eq.~\eqref{eq:SDE_bull} in the main text), mechanical driving from the molecular motors dominates. As shown in Fig.~3b, observing $x^1$ allows access to approximately 50\% of the total entropy production.}

\subsection{Red blood cell model}

\begin{figure*}[t!]
 	\centering
\includegraphics[width=1\textwidth]{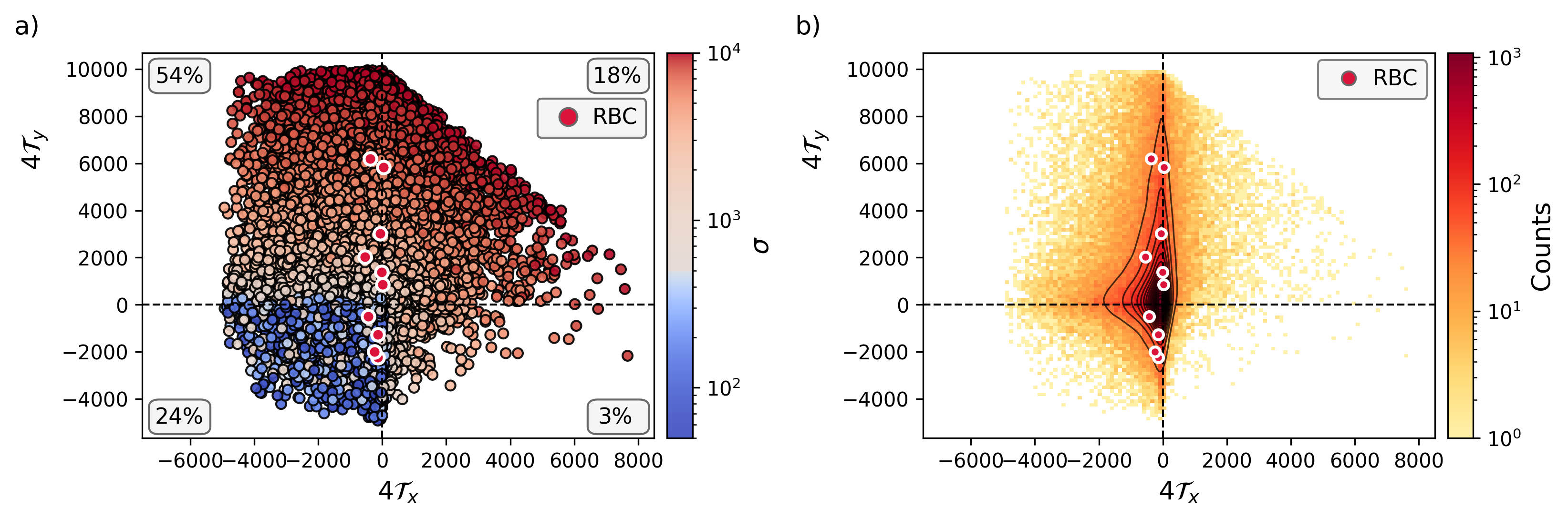}
	\caption{a) Realizations of the RBC model in Eq.~\eqref{SDE_RBC_resp} with $\sigma \le 10^4\,k_{\rm B}/s$ and with randomly and homogeneously sampled parameters in the following intervals: $\mu_x \in (5\cdot 10^3,\, 4 \cdot 5\cdot 10^4)$, $\mu_y \in (5\cdot 10^3,\, 4 \cdot 5\cdot 10^4)$, $k_x \in (10^{-3},\, 10^{-1})$, $k_y \in (10^{-3},\, 10^{-1})$, $k_{\rm int} \in (10^{-3},\, 10^{-1})$, $\epsilon \in (0.5, 20)$ and $\tau \in (5\cdot 10^{-3},\, 5\cdot 10^{-1})$. Red dots represent experimental estimates for selected highly active RBCs in Ref.\cite{VSR}. Percentage in all quadrants reflect the number of realizations falling in that quadrant. b) Histogram of the same realizations shown in panel~a), with overlaid contour lines obtained via a kernel density estimate. Red dots are the same estimates for RBCs shown in panel a).}  
	\label{Fig_resp_2}
\end{figure*}

{To further investigate how detectability depends on the structure of the system and its couplings, we now turn to the red blood cell (RBC) model introduced in Ref.~\cite{VSR}. The dynamics involve three degrees of freedom $(x_t, y_t, f_t)$ and are described by the following set of stochastic differential equations:
\begin{equation}\label{SDE_RBC_resp}
\begin{split}
    \dot{x}_t &= \mu_{x} \left( -k_x x_t + k_{\rm int} y_t \right) + \sqrt{2 D_x}\, \xi^x_t \,,\\
    \dot{y}_t &= \mu_{y} \left( -k_y y_t + k_{\rm int} x_t +f_t \right) + \sqrt{2 D_y}\, \xi^y_t \,,\\
    \dot{f}_t &= -f_t/\tau + \sqrt{2 \epsilon^2/\tau} \,\xi^f_t \,.
\end{split}
\end{equation}
The model can be analytically solved, see \cite{VSR} for details. Here, nonequilibrium arises from the active forcing $f_t$, while only the degree of freedom $x_t$ is accessible experimentally. To assess whether nonequilibrium signatures can be detected from the observation of a single variable, we computed $\sigma$ and the traffic components $\mathcal{T}_x$ and $\mathcal{T}_y$ over a high number of parameter sets satisfying physical constraints ($k_x k_y - k_{\rm int}^2 > 0$), and filtered for $\sigma \le 10^4\,k_{\rm B}/s$, a biologically reasonable scale. Results are shown in Fig.~\ref{Fig_resp_2}, with $4\mathcal{T}_x$ and $4\mathcal{T}_y$ on the horizontal and vertical axes, respectively, and $\sigma$ encoded in color in panel a). Panel b) complements this by showing a two-dimensional histogram of the same realizations, where each bin reflects how many simulations fall within a given region of the $(4\mathcal{T}_x, 4\mathcal{T}_y)$ space. Overlaid contour lines obtained via kernel density estimation highlight areas of high density, revealing that realizations tend to cluster around the vertical axis, where $\mathcal{T}_y$ is large and $\mathcal{T}_x$ is small or negative. Notably, the red dots, corresponding to experimental estimates for highly active RBCs with $\sigma > 200\,k_{\rm B}/s$ whose experimental parameters were estimated in \cite{VSR}, appear in this region of high density, suggesting that these experiments fall into a parameter regime where dissipation is largely carried by $y$, rendering detection from $x$ alone more difficult. For small $\sigma$, and for our particular choice of sampled parameters, both traffic components are negative in about 24\% of the cases, precluding detection from either $x$ or $y$. As $\sigma$ increases, $\mathcal{T}_y$ rises rapidly and usually accounts for most of the dissipation, while $\mathcal{T}_x$ remains smaller and only becomes positive in a subset of cases. The most frequent regime (54\%), corresponding to $\mathcal{T}_y > 0$ and $\mathcal{T}_x < 0$, was the one encountered in our previous work~\cite{VSR}, where $\sigma \sim 10^3\,k_{\rm B}/s$ and was insufficient to make $\mathcal{T}_x$ positive (see again red dots in Fig.~\ref{Fig_resp_2}a). This explains the need for a model-based approach in \cite{VSR}. Moreover, as $\sigma$ increases, the likelihood of $\mathcal{T}_x > 0$ also rises, reaching 18\% in our sampled sets. These trends suggest that traffic-based detectability improves when the observed degree of freedom is closer to the energy source. It is worth noting that no other model-free method can currently be applied when only $x$ is observed.}

{Figure~\ref{Fig_resp_2} also shows that access to both $x$ and $y$ would allow for partial inference of $\sigma$ in most cases, in particular in the RBC experiments. In this context, one could also apply the TUR by optimizing over a class of currents. A body of works~\cite{manikandan2020inferring,otsubo2020estimating,manikandan2021quantitative} has shown that the short-time TUR is particularly effective, especially when the observed current is the entropy production itself. Since that current is not directly accessible, we focus on generalized circulating currents of the form:
\begin{equation}
    J(x_t,y_t) = \int_{0}^{t} \mathrm{d}s\left( x_s\circ \dot{y}_s - \alpha \, y_s\circ \dot{x}_s \right),
\end{equation}
parametrized by a scalar $\alpha$. In Section~\ref{current_avg_var} we find that their mean and variance at short times are:
\begin{equation}\label{current_avg_and_var}
\begin{split}
\langle J_{dt} \rangle_{\alpha} &\approx (1+\alpha) \dot{C}_{xy}(0) dt,\\
\langle \Delta J^2_{dt} \rangle_{\alpha} &\approx  k_{\rm B} T \left(\mu_y C_{xx}(0) + \alpha^2 \mu_x C_{yy}(0)\right)dt,
\end{split}
\end{equation}
and since $\Sigma_{\rm tot} \approx \sigma dt$, we obtain the bound:
\begin{equation}
\frac{2 \langle J_{dt} \rangle_{\alpha}^{2}}{\langle \Delta J^2_{dt} \rangle_{\alpha}}\le \sigma dt \implies \frac{ 2(1+\alpha)^2 \dot{C}_{xy}^{\,2}(0)}{k_{\rm B} T (\mu_y C_{xx}(0) + \alpha^2 \mu_x C_{yy}(0))} \le \sigma.
\end{equation}
Defining:
\begin{equation}\label{sigma_TUR}
\sigma_{\rm TUR} = \max_\alpha \left[ \frac{ 2(1+\alpha)^2 \dot{C}_{xy}^{\, 2}(0)}{k_{\rm B} T (\mu_y C_{xx}(0) + \alpha^2 \mu_x C_{yy}(0))} \right],
\end{equation}
we can compare this TUR-based estimator to the kinetic bound from the main text:
\begin{equation}
\sigma_{\mathcal{S}} = \sum_{i \in \mathcal{S}} \max(4 \mathcal{T}_i, 0) \le \sigma.
\end{equation}
}

\begin{figure*}[t!]
 	\centering
\includegraphics[width=1\textwidth]{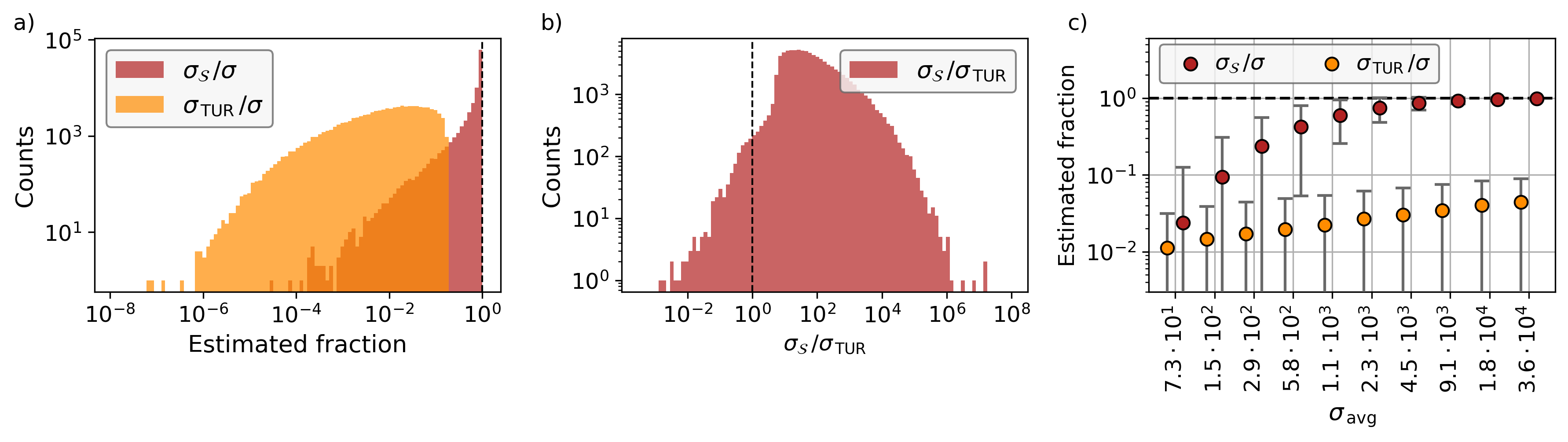}
	\caption{
Comparison between TUR-based and traffic-based estimators of entropy production for the RBC model. 
a) Histogram of the normalized estimators $\sigma_{\mathcal{S}}/\sigma$ (red) and $\sigma_{\rm TUR}/\sigma$ (orange) across random parameter sets as illustrated in the caption of Fig.~\ref{Fig_resp_2} and with $\sigma \le 5\cdot10^4\,k_{\rm B}/s$. Only values $\sigma_{\mathcal{S}}>0$ are shown due to log-scaling. b) Histogram of the ratio $\sigma_{\mathcal{S}}/\sigma_{\rm TUR}$ for the same dataset, illustrating the superior performance of the kinetic bound in the majority of cases. c) Average values of $\sigma_{\mathcal{S}}/\sigma$ (red) and $\sigma_{\rm TUR}/\sigma$ (orange), binned by increasing $\sigma$ (logarithmic bins), with error bars representing standard deviations. The kinetic bound saturates toward unity as $\sigma$ increases, while the TUR estimator remains low and does not exceed 5\% on average. }
\label{Fig_resp_3}
\end{figure*}

{Results are summarized in Fig.~\ref{Fig_resp_3}, restricted to cases with $\sigma \le 5 \cdot 10^4\,k_{\rm B}/s$, and where $\sigma_{\rm \, TUR}$ has been obtained by taking the maximum value of the right-hand side of Eq.~\eqref{sigma_TUR} over $10^3$ values of  $\alpha \in (10^{-2},10^2)$, uniformly sampled in log-space. In Figs.~\ref{Fig_resp_3}a and~\ref{Fig_resp_3}b, histograms for $\sigma_{\mathcal{S}}$ are shown for $\sigma_{\mathcal{S}} \neq 0$ due to log-scaling in the x-axis of the plot. These zero cases represent 18\% of the total but become rare at higher $\sigma$: 5\% for $\sigma \ge 500 \,k_{\rm B}/s$ and 2\% for $\sigma \ge 1000\,k_{\rm B}/s$. Fig.~\ref{Fig_resp_3}a compares $\sigma_{\rm TUR}/\sigma$ (orange) and $\sigma_{\mathcal{S}}/\sigma$ (red). On average, the kinetic bound is far more informative: $\langle \sigma_{\rm TUR}/\sigma \rangle \approx 0.02$, versus $\langle \sigma_{\mathcal{S}}/\sigma \rangle \approx 0.7$ (including zeros). Fig.~\ref{Fig_resp_3}b shows the ratio $\sigma_{\mathcal{S}} / \sigma_{\rm TUR}$, confirming that kinetic estimates outperform TUR-based bounds by several orders of magnitude in most cases. Finally, Fig.~\ref{Fig_resp_3}c shows the average performance as a function of increasing $\sigma$ (x-axis labels show the average value of $\sigma$ of all realizations contributing to the values plotted on the y-axis): the kinetic bound saturates at high $\sigma$, while the TUR-based estimate remains capped at around 5\%. This saturation arises because the inflow rate becomes negligible at large $\sigma$, with traffic capturing most of the dissipation, as shown in Section \ref{InfoG}.}

\subsubsection{Short-time statistics of generalized currents}\label{current_avg_var}

{In this section, we compute the short-time average and variance of the following family of generalized currents:
\begin{equation}\label{J_current}
    J(x_t,y_t) = \int_{0}^t \mathrm{d} s \,j_s = \int_{0}^{t} \mathrm{d}s\left( x_s\circ \dot{y}_s - \alpha \, y_s\circ \dot{x}_s \right) \, .
\end{equation}
We begin with the average. At steady state, it reads:
\begin{equation}
\langle J_{dt} \rangle_{\alpha} = \langle j \rangle dt= \left( \langle x  \dot{y} \rangle - \alpha \langle y  \dot{x} \rangle \right)dt \, .
\end{equation}
To compute $\langle x \dot{y} \rangle$, we express the Stratonovich product explicitly:
\begin{equation}
    \langle x \circ \dot{y} \rangle  = \frac{1}{2dt}\langle \left( x_{t+dt} + x_{t}  \right)\left( y_{t+dt}-y_{t} \right) \rangle = \frac{1}{2 dt} \left(  C_{yx}(dt) -C_{xy}(dt)\right),
\end{equation}
where we used that $C_{AB}(\tau) = \langle A_{t+\tau}B_t \rangle$ and $C_{xy}(0) = C_{yx}(0)$ in steady state. By performing a Taylor expansion at first order:
\begin{equation}
    C_{xy}(dt) \approx C_{xy}(0) + \dot{C}_{xy}(0)\, dt, \quad C_{yx}(dt) \approx C_{yx}(0) + \dot{C}_{yx}(0)\, dt\, .
\end{equation}
we get
\begin{equation}
    \langle x \dot{y} \rangle =  \frac{1}{2}  \left( \dot{C}_{yx}(0) -\dot{C}_{xy}(0) \right).
\end{equation}
From Section~\ref{proof_D}, we recall that $\pmb{D} = - \dot{\pmb{C}}^{\, \rm S}_x(0)$, implying that $\dot{C}_{xy}(0) + \dot{C}_{yx}(0) = -2D_{xy} = 0$, hence $\dot{C}_{yx}(0) = -\dot{C}_{xy}(0)$. Therefore:
\begin{equation}\label{x_y_dot}
    \langle x \dot{y} \rangle = \dot{C}_{yx}(0) = -\langle y  \dot{x} \rangle \,.
\end{equation}
This yields a compact expression for the current:
\begin{equation}
\langle J_{dt} \rangle_{\alpha} = (1+\alpha)\dot{C}_{xy}(0)dt,
\end{equation}
which proves the first line of Eq.~\eqref{current_avg_and_var}.}

{We now compute the variance of $J_t$:
\begin{equation}
    \langle \Delta J^2_{t} \rangle_{\alpha} = \int_{0}^{t}\mathrm{d}t^{\prime}\int_{0}^{t}\mathrm{d}t^{\prime\prime} \left(\langle j_{t^{\prime}} j_{t^{\prime\prime}} \rangle - \langle j_{t^{\prime}} \rangle \langle j_{t^{\prime\prime}} \rangle \right) = 2 \int_{0}^{t}\mathrm{d}t^{\prime}\int_{0}^{t^{\prime}}\mathrm{d}t^{\prime\prime} C_{j}(t^{\prime\prime}),
\end{equation}
which at leading order for short $t = dt$ becomes:
\begin{equation}\label{J_C_j}
    \langle \Delta J^2_{dt} \rangle_{\alpha} \approx dt^{2} C_{j}(0).
\end{equation}
To compute $C_j(0)$, we start from the expression of $j_t$:
\begin{equation}
    j_t = x_t \circ \dot{y}_t - \alpha y_t \circ \dot{x}_t,
\end{equation}
which gives:
\begin{equation}\label{Cj_0_step1}
    C_j(0) = \langle x^2 \dot{y}^2 \rangle + \alpha^2 \langle y^2 \dot{x}^2 \rangle - 2\alpha \langle x y \dot{x} \dot{y} \rangle - \langle x \dot{y} \rangle^2 - \alpha^2 \langle y \dot{x} \rangle^2 + 2\alpha \langle x \dot{y} \rangle \langle y \dot{x} \rangle.
\end{equation}
Because the system is linear and in a steady-state, all variables are Gaussian. We can therefore apply Wick’s theorem:
\begin{equation}
\langle abcd \rangle = \langle ab \rangle \langle cd \rangle + \langle ac \rangle \langle bd \rangle + \langle ad \rangle \langle bc \rangle,
\end{equation}
to compute:
\begin{equation}
\begin{split}
    \langle x^2 \dot{y}^2 \rangle &= \langle x^2 \rangle \langle \dot{y}^2 \rangle + 2\langle x \dot{y} \rangle^2, \\
    \langle y^2 \dot{x}^2 \rangle &= \langle y^2 \rangle \langle \dot{x}^2 \rangle + 2\langle y \dot{x} \rangle^2, \\
    \langle x y \dot{x} \dot{y} \rangle &= \langle x y \rangle \langle \dot{x} \dot{y} \rangle + \langle x \dot{x} \rangle \langle y \dot{y} \rangle + \langle x \dot{y} \rangle \langle y \dot{x} \rangle.
\end{split}
\end{equation}
Using that $\langle x \dot{x} \rangle = \langle y \dot{y} \rangle = 0$, this leads to:
\begin{equation}\label{Cj_0_step12}
    C_j(0) = \langle x^2 \rangle \langle \dot{y}^2 \rangle + \alpha^2 \langle y^2 \rangle \langle \dot{x}^2 \rangle + \langle x \dot{y} \rangle^2 + \alpha^2 \langle y \dot{x} \rangle^2 - 2\alpha \langle x y \rangle \langle \dot{x} \dot{y} \rangle.
\end{equation}
Now we evaluate the remaining second moments. For example:
\begin{equation}
\begin{split}
    \langle \dot{x}\dot{y} \rangle = \langle \dot{x}\circ\dot{y} \rangle &= \frac{1}{2}\Big\langle  \left( \frac{x_{t+2dt} - x_{t+dt}}{dt} +  \frac{x_{t+dt} - x_{t}}{dt}\right)\left( \frac{y_{t+dt}-y_{t}}{dt} \right) \Big\rangle\\
    &= \frac{1}{2dt^2}\langle \left( x_{t+2dt}  - x_{t}\right)\left( y_{t+dt}-y_{t} \right)\rangle\\
    &= \frac{1}{2dt^2} \left( C_{xy}(dt) + C_{xy}(0) -C_{xy}(2dt) -C_{yx}(dt) \right) \\
    &= -\frac{1}{2dt} \left( \dotnice{C}_{xy}(0) + \dotnice{C}_{yx}(0) \right) = \frac{D_{xy}}{dt} = 0
\end{split}
\end{equation}
where we used again that $\pmb{D} = - \dot{\pmb{C}}^{\rm S}_x(0)$. Similarly, the variances of $\dot{x}$ and $\dot{y}$ are:
\begin{align}
    \langle \dot{x}^2 \rangle = \frac{D_{xx}}{dt} = \frac{k_{\rm B} T \mu_x}{dt}, \quad \langle \dot{y}^2 \rangle = \frac{D_{yy}}{dt} = \frac{k_{\rm B} T \mu_y}{dt}.
\end{align}
Note that these terms scale as $1/dt$, while $\langle x \dot{y} \rangle$ and $\langle y \dot{x} \rangle$ are $\mathcal{O}(1)$ (see Eq.~\eqref{x_y_dot}). Therefore, at leading order, Eqs.~\eqref{J_C_j} and \eqref{Cj_0_step12} yield:
\begin{equation}
    \langle \Delta J^2_{dt} \rangle_{\alpha} \approx dt^{2} C_j(0) \approx k_{\rm B} T \left( \mu_y C_{xx}(0) + \alpha^2 \mu_x C_{yy}(0) \right) dt \,,
\end{equation}
which completes the proof.}

\section{Hair-cell model}\label{hair_cell_model}

The spontaneous oscillations of hair bundles in the auditory organs of bullfrogs are driven by the interplay of mechanosensitive ion channels, molecular motor activity, and calcium feedback mechanisms. These dynamics, derived and discussed in \cite{hair0,hair1,hair2,hair3,roldan2021quantifying}, are effectively captured by a model with two degrees of freedom: the bundle position $x^1$ and the center of mass of the molecular motors $x^2$. Following \cite{roldan2021quantifying}, in this section, we provide a brief overview of the key equations governing the system's evolution. We outline the roles of potential $V(x^1, x^2)$, active force $F_t^{\rm act}$, and effective temperature $T^{\rm eff}$, highlighting their contributions to the nonequilibrium nature of the system.

The system's equations are:
\begin{equation} 
\begin{split} 
\dot{x}^1_t &= -\mu_1 \partial_{x^1}V_{t}+\sqrt{2k_{\rm B}T\mu_1}\,\xi^1_t \, ,\\[3pt] 
\dot{x}^2_t &= -\mu_2 \partial_{x^2}V_{t}-\mu_2F_{t}^{\rm act}+\sqrt{2k_{\rm B}T^{\rm eff}\mu_2}\,\xi^1_t \, ,
\end{split} 
\end{equation}
where $\mu_1$ and $\mu_2$ are mobility coefficients and where the potential $V_t = V(x^1_t, x^2_t)$ accounts for elastic forces and mechanosensitive ion channels. The explicit form of $V(x^1, x^2)$ is given by:
\begin{equation}
    V(x^1, x^2) = \frac{k_{gs}\left( x^1-x^2\right)^2 + k_{sp}(x^1)^2}{2} 
    - Nk_{\rm B} T \ln \left[ \exp\left( \frac{k_{gs}D\left( x^1-x^2\right)}{N k_{\rm B} T }\right)+A \right],
\end{equation}
where $k_{gs}$ and $k_{sp}$ are stiffness coefficients, $D$ is the gating swing of a transduction channel, and 
\begin{equation}
    A = \exp\left[\left(\Delta G + \frac{k_{gs}D^2}{2N}\right)/k_{\rm B} T\right]\, .
\end{equation}
Here, $\Delta G$ represents the energy difference between the open and closed states of the channels, and $N$ is the number of transduction elements. The nonequilibrium behavior is driven by molecular motor activity, which is encoded in the effective temperature $T^{\rm eff}$ and the non-conservative force:
\begin{equation}
    F_{t}^{\rm act} = F^{\rm max}(1 - SP_0(x^1_t, x^2_t)) \, .
\end{equation}
The parameter $S$ quantifies calcium-mediated feedback on the motor force, and the open probability of the transduction channels is given by:
\begin{equation}
    P_{0}(x^1,x^2) = \frac{1}{1+A \exp\left(-k_{gs}D(x^1-x^2)/(N k_{\rm B} T)\right)}.
\end{equation}
This expression for $P_0(x^1, x^2)$ represents the open probability of a two-state equilibrium model of a channel, where the difference in free energy between the open and closed states depends linearly on the distance $x^1-x^2$.

\begin{figure*}[t!]
 	\centering
\includegraphics[width=1\textwidth]{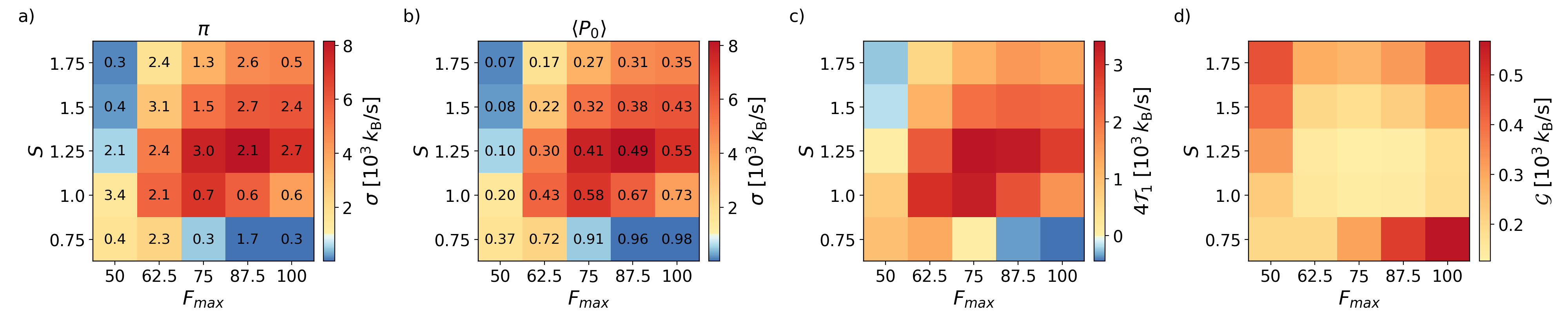}
\caption{a) Heatmap of $\sigma$ as a function of the dissipative parameters $F^{\rm max}$ and $S$. The accuracy $\pi(\overline{\sigma})$ for each combination of $F^{\rm max}$ and $S$ is displayed above the corresponding values in the heatmap. b) Same heatmap of $\sigma$, now annotated with the average ion channel opening probability $\langle P_0 \rangle$ displayed above each value.
c) Heatmap of $4\traffic_{1}$, exhibiting a similar color pattern to panel a).
d) Heatmap of the inflow rate $\inflow$, showing minimal variability across all samples compared to panels a) and b), as well as significantly smaller values relative to the maximum $\sigma$ observed in panel a).}
	\label{Fig_supp_bull}
\end{figure*}
\begin{figure*}[t!]
 	\centering
\includegraphics[width=0.35\textwidth]{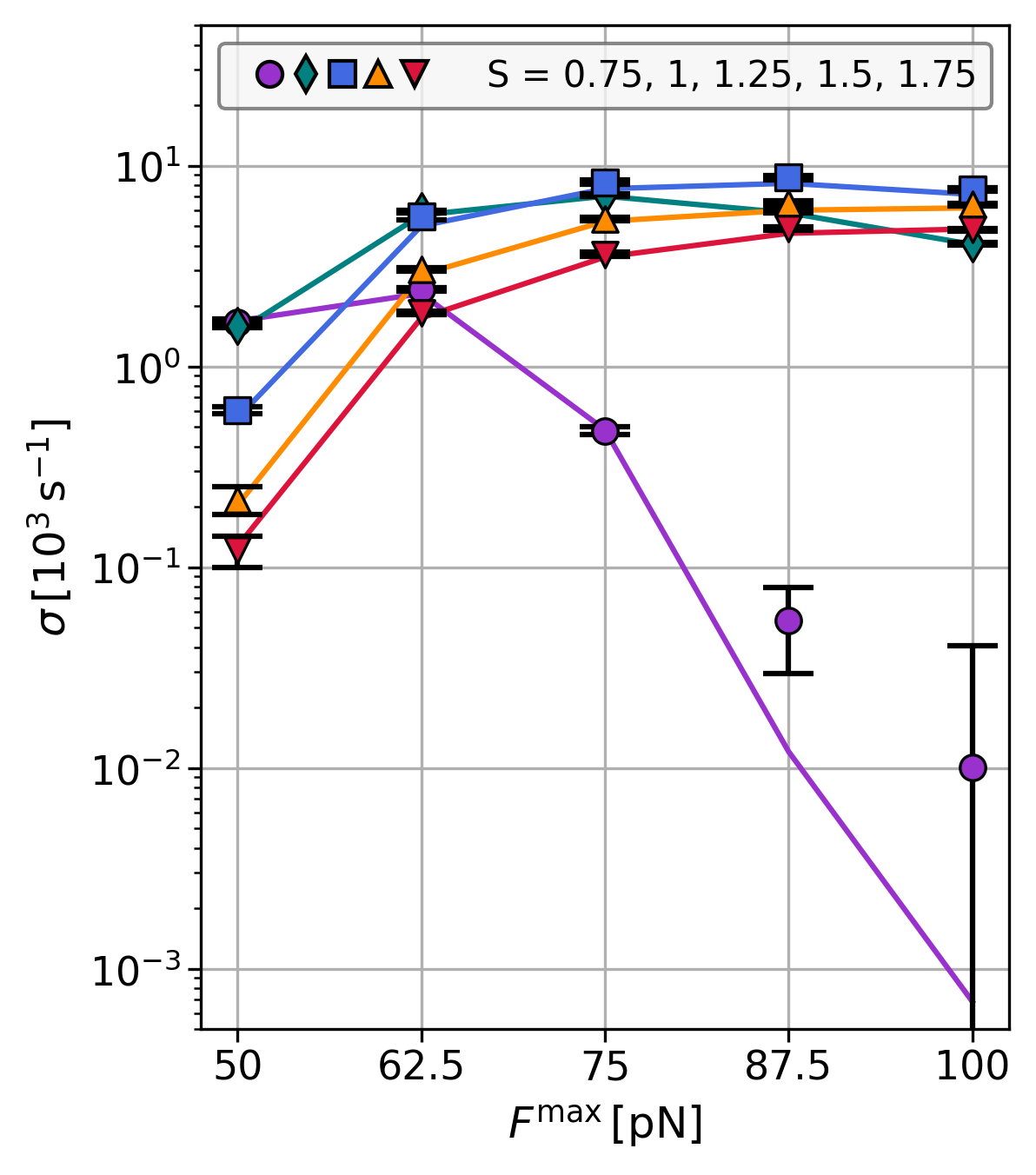}
\caption{Panel from Fig.~2d in main text, displayed on a logarithmic scale to highlight the full range of entropy production values and better illustrate the performance of the estimation across several orders of magnitude. {fix ylabel}}
	\label{Bull_perf}
\end{figure*}

In the main text we discussed the application of our method to simulated traces with fixed parameters. These are set to are set to: $ \mu_1 = 357\, [{\rm nm/(pN s)}]$ , $ \mu_2 = 100 \,[{\rm nm/(pN s)}]$, $k_{gs} = 0.75 \,[{\rm pN/nm}] $, $k_{sp} =0.6p\,[{\rm N/nm}]$, $D=61\,[{\rm nm}]$, $N=50$ and $\Delta G=10k_{\rm B}T$. Furthermore, in Fig.\ref{Fig_supp_bull}, we present additional results that illustrate the behavior of the system as a function of $F^{\rm max}$ and $S$. Panel a) shows a heat map of $\sigma$, with accuracy $\pi(\overline{\sigma})$  displayed above each combination of parameters to quantify the reliability of our estimates. Panel b) presents the heatmap of $4\traffic_{1}$, displaying a color pattern that closely resembles that of panel a). This similarity demonstrates how dissipation is effectively captured and encoded within the traffic components. Finally, panel c) illustrates the inflow rate $\inflow$, which exhibits markedly lower variability compared to $\sigma$ and $4\traffic_{1}$ across all combinations of parameters. Furthermore, the values of $\inflow$ are significantly lower than the maximum $\sigma$ observed in panel a), reinforcing the conclusion that information about the dissipative processes is primarily encoded in the traffic $\traffic$, rather than in the inflow $\inflow$.

\section{Data analysis}\label{Data_analysis}

The estimation process involves two essential steps to derive the entropy production rate $\sigma$ from a single stochastic trace using Eq.\eqref{main_eq}. First, the short-time first and second derivatives of the correlation functions, $\dotnice{C}^{\, ij}_{x}(t)$ and $\ddotnice{C}^{\, ij}_{x}(t)$, are computed. As Appendix \ref{proof_D} outlines, the first derivative determines the diffusion matrix $D$, which is needed to calculate both terms in the main equation \eqref{main_eq}. The second step calculates the inflow rate $\inflow$, which instead focuses on calculating the correlation matrix of the effective forces $-\nabla \phi(x)$, $\pmb{C}_{\nabla \phi}(0)$.

\subsection{Estimation of derivatives}

For the first task, each component of the position correlation matrix $\pmb{C}_{x}(t)$ is fitted with a sum of an appropriate, system-dependent number of exponentials:
\begin{equation}
(\pmb{C}_x^{\,\text{fit}}(t))^{ij} = \sum_k A_k e^{-t / \tau_k},
\end{equation}
where $A_k$ are the amplitudes and $\tau_k$ the system's typical timescales. From this representation, the short-time derivatives can be derived analytically:
\begin{equation}\label{der_expr_pars}
\dotnice{C}^{\, ij}_{x}(t) = -\sum_k \frac{A_k}{\tau_k}\,, \quad \ddotnice{C}^{\, ij}_{x}(t) = \sum_k \frac{A_k}{\tau_k^2}.
\end{equation}
In practice, it is sufficient to fit only the components of the symmetric correlation matrix, as these are the only terms required in the relevant formulas. This approach eliminates the need for numerical differentiation, which is particularly susceptible to noise in stochastic data. For clarity, the $ij$ component indices will also be omitted from the formulas hereafter.

\begin{figure*}[t!]
 	\centering
    \includegraphics[width=0.9\textwidth]{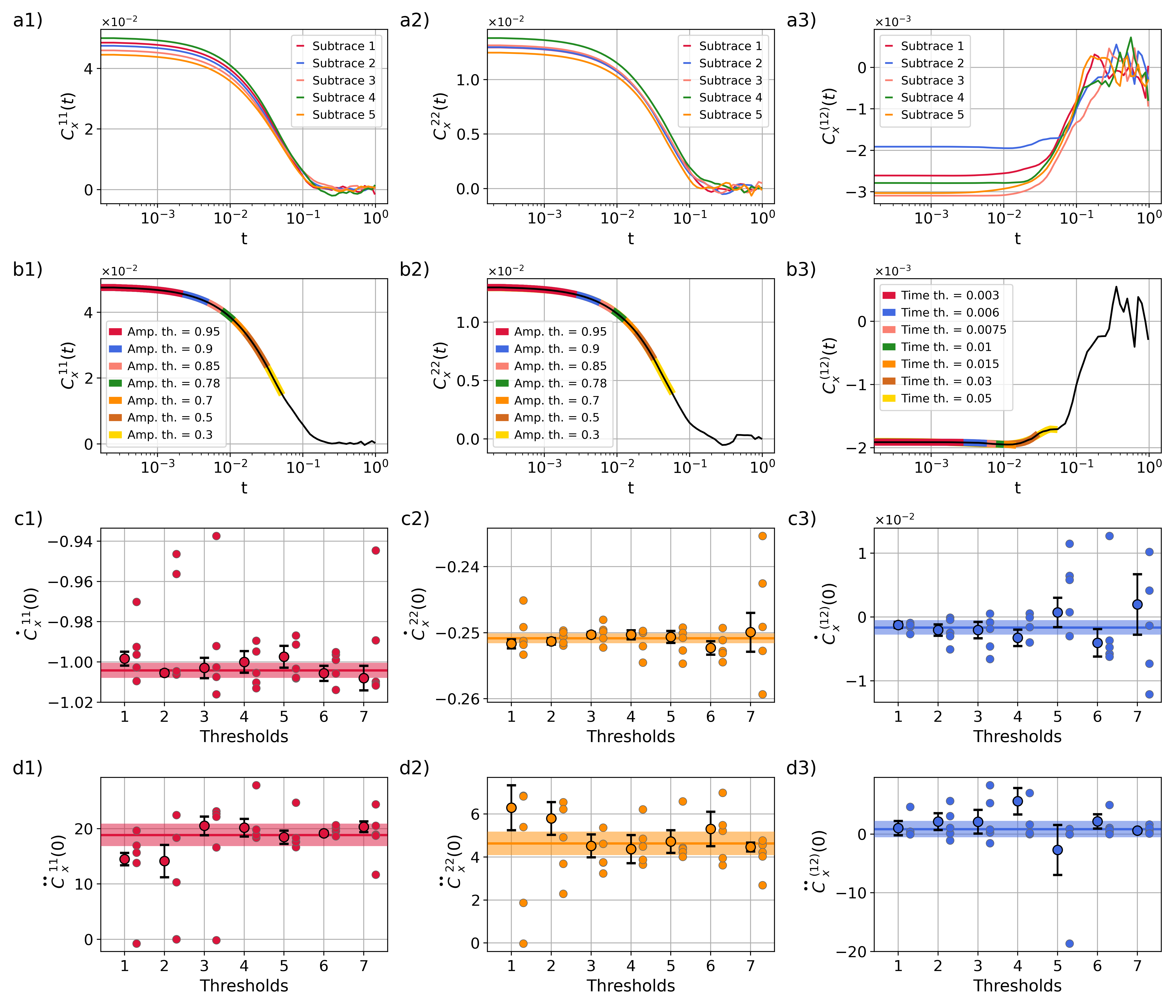}
	\caption{Derivatives estimation for linear model with $\alpha_A = -0.5$ and $\alpha_T=0.25$. a1,a2,a3) Components of the symmetrized correlation matrix $\pmb{C}^{\rm \, S}_{x}(t)$ for all subtraces. b1,b2,b3) Fits for different thresholds for one particular subtrace. c1,c2,c3) Results of the estimation for the components $\dot{\pmb{C}}^{\rm \, S}_{x}(t)$ across various subtraces and thresholds.
    d1,d2,d3) Estimation results for the components $\ddot{\pmb{C}}^{\rm \, S}_{x}(t)$ across the same subtraces and thresholds. }
	\label{FIG_der_lin}
\end{figure*}

\begin{figure*}[t!]
 	\centering
	\includegraphics[width=0.9\textwidth]{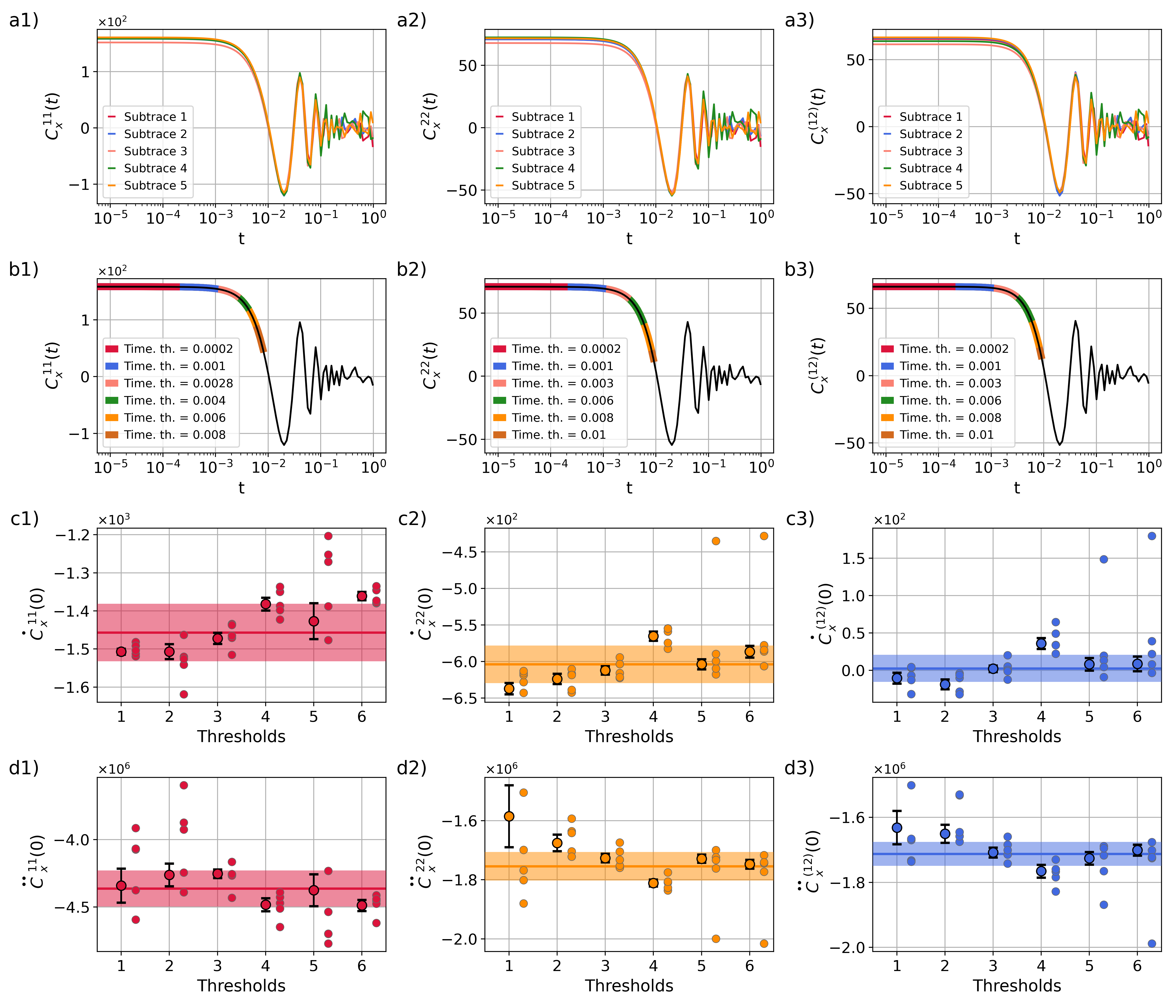}
	\caption{Derivatives estimation for Hair-cell model with $F^{\rm max} = 62.5$ and $S=1$.  a1,a2,a3) Components of the symmetrized correlation matrix $\pmb{C}^{\rm \, S}_{x}(t)$ for all subtraces. b1,b2,b3) Fits for different thresholds for one particular subtrace. c1,c2,c3) Results of the estimation for the components $\dot{\pmb{C}}^{\rm \, S}_{x}(t)$ across various subtraces and thresholds.
    d1,d2,d3) Estimation results for the components $\ddot{\pmb{C}}^{\rm \, S}_{x}(t)$ across the same subtraces and thresholds. }
	\label{FIG_der_bull}
\end{figure*}

The stochastic trace is split into $n$ subtraces to ensure robustness, with each subtrace analyzed independently. For each subtrace, position correlation functions are numerically calculated over a predefined time window, see first row in Figures \ref{FIG_der_lin} and \ref{FIG_der_bull}. A threshold-based approach is applied to select data points for fitting. These thresholds are defined as either a fraction of the initial amplitude of the correlation function or as a cut-off in time, depending on the shape of the correlation functions.

The fitting process determines the parameters $A_k$ and $\tau_k$ for each subtrace and threshold. The fit results are shown in the second row of Figures \ref{FIG_der_lin} and \ref{FIG_der_bull} for a selected subtrace and all chosen thresholds, where the correlation functions are plotted along with their exponential fits, illustrating the data and model agreement. The quality of the fits is then evaluated using a residual metric, denoted by $\chi^2$, defined as:
\begin{equation}
\chi^2 = \frac{1}{\nu} \sum_{i=1}^N \left(C_x^{\,\text{fit}}(t_i) - C_x(t_i)\right)^2,
\end{equation}
where $N$ is the number of data points in time, $\nu = N - p$ is the number of degrees of freedom, $p$ is the number of fitting parameters, $C_x(t_i)$ is the observed correlation at time $t_i$, and $C_x^{\, \text{fit}}(t_i)$ is the corresponding fitted value.

In many applications, residual metrics like $\chi^2$ are normalized by dividing the squared residuals by $C_x(t_i)$ to weight the contribution of each point proportionally to its magnitude. However, this normalization is not used here because it leads to instability when $C_x(t_i)$ approaches zero. By avoiding this normalization, the residual metric ensures a robust evaluation of fit quality without overweighting regions where correlation functions are close to zero. The inclusion of $1/\nu$ adjusts for the degrees of freedom in the fit but does not affect the computation of averages or variances in subsequent steps, as we take $\nu$ to be the same for all subtraces at a given threshold. The squared inverse of this metric, \(w = \chi^{-2}\), is then used as the weight for each fit at a fixed threshold, ensuring that more reliable fits contribute more significantly to the initial estimates of \(\dotnice{C}_x(0)\) and \(\ddotnice{C}_x(0)\).
\begin{figure*}[t!]
 	\centering
	\includegraphics[width=1\textwidth]{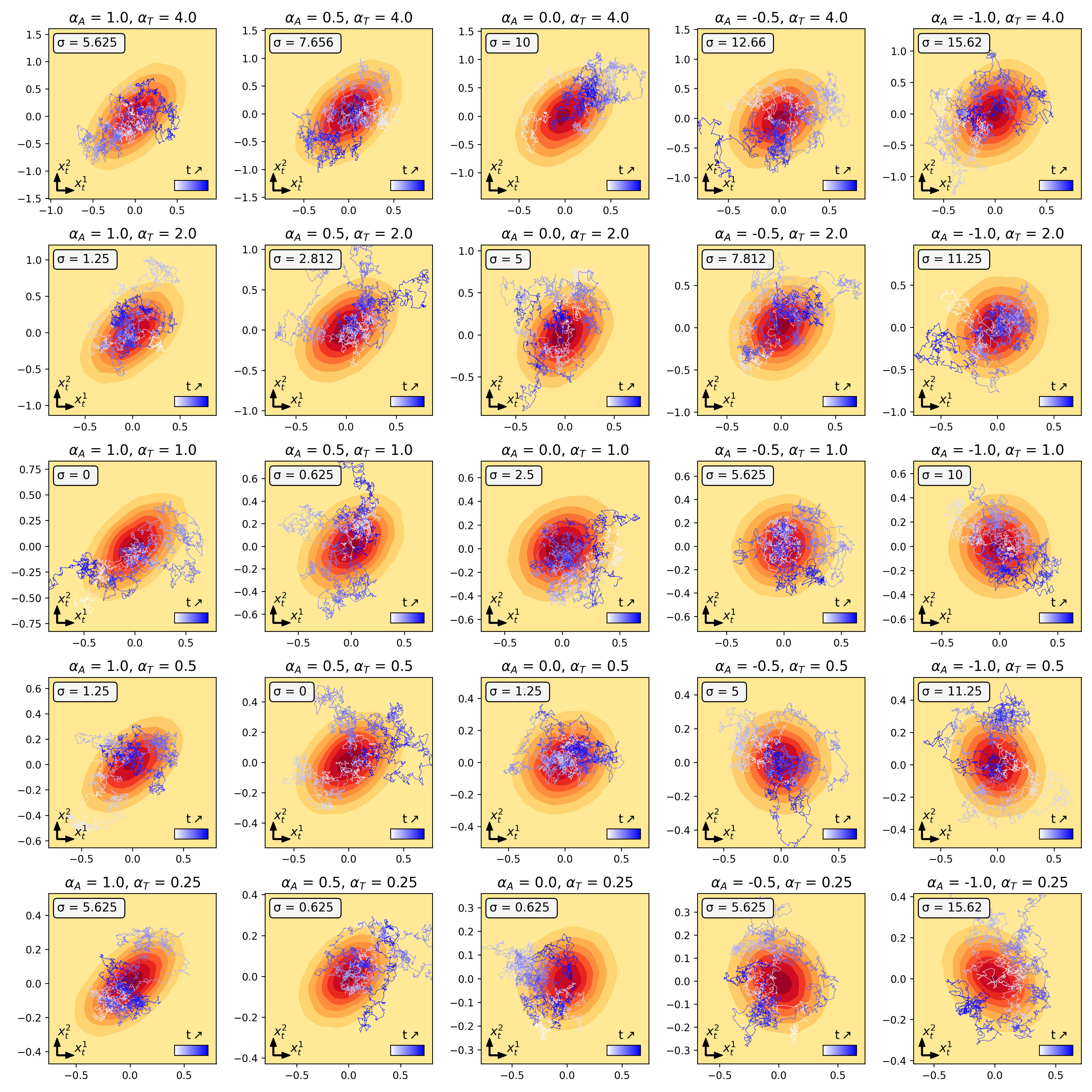}
	\caption{$p(\pmb{x}_t)$ of all analyzed traces corresponding to the linear model, represented as a contour map. Regions with a deeper red color indicate higher values of $p(\pmb{x}_t)$. A segment of a stochastic trace is overlaid on the contour map to illustrate its dynamics within the probability landscape.}
	\label{FIG_all_pdf_lin}
\end{figure*}
As a result, at each threshold, the derivative estimates across all subtraces are combined using a weighted averaging procedure. Specifically, the weighted mean for a derivative $X \in \{\dotnice{C}_x(0), \ddotnice{C}_x(0)\}$, estimated through Eq.~\eqref{der_expr_pars} with the fitted parameters, is calculated as:
\begin{equation}
\langle X \rangle_{\text{th}} = \frac{\sum_{i} w_i X_i}{\sum_{i} w_i}\,,
\end{equation}
where $X_i$ is the estimate for a particular subtrace, and $w_i = 1 / \chi_i^2$ is the associated weight derived from the residual metric. The variance of the weighted mean is calculated as:
\begin{equation}
\sigma^2_{X, \text{th}} = \frac{\sum_{i} w_i \left(X_i - \langle X \rangle_{\text{th}}\right)^2}{\sum_{i} w_i}\,.
\end{equation}
We further combine these values to produce the final estimate of $\dotnice{C}_x(0)$ and $\ddotnice{C}_x(0)$. This is achieved using the same weighted averaging procedure, treating the threshold-specific means as individual estimates and their variances as the basis for weights. The final weighted mean is computed as:
\begin{equation}
\langle X \rangle = \frac{\sum_{\text{th}} w_{\text{th}} \langle X \rangle_{\text{th}}}{\sum_{\text{th}} w_{\text{th}}}\,,
\end{equation}
where $w_{\text{th}} = \sigma_{X,\textrm{th}}^{-2}$ is the weight derived from the threshold-specific variance. The variance of the weighted mean is calculated analogously:
\begin{equation}
\sigma_X^2 = \frac{\sum_{\text{th}} w_{\text{th}} \left(\langle X \rangle_{\text{th}} - \langle X \rangle\right)^2}{\sum_{\text{th}} w_{\text{th}}}\,,
\end{equation}
with the standard error given by
$\sigma_{\,\overline{X}} = \sqrt{\sigma_X^2 / (N_{\rm th} - 1)}$,
where $N_{\rm th}$ is the number of chosen thresholds.

The results of this two-level averaging procedure are shown in the third and fourth row of Figures \ref{FIG_der_lin} and \ref{FIG_der_bull}. The derivative estimates for all thresholds are also displayed, with weighted averages and their uncertainties highlighted to confirm the reliability and consistency of the approach.

By splitting stochastic traces into $n$ subtraces and aggregating results first across subtraces and then across thresholds, this two-level averaging procedure provides robust and reliable estimates of $\dotnice{\pmb{C}}_x(0)$ and $\ddotnice{\pmb{C}}_x(0)$. By finally using that $\pmb{D} = -\dotnice{\pmb{C}}_x(0)$, we can readily calculate the ${\rm Tr}\!\left[\pmb{D}^{\rm-1}\ddotnice{\pmb{C}}_{x}(0) \right]$ term in \eqref{main_eq} with estimate errors calculated with the error propagation formula.

\subsection{Estimation of effective forces}
\begin{figure*}[t!]
 	\centering
	\includegraphics[width=1\textwidth]{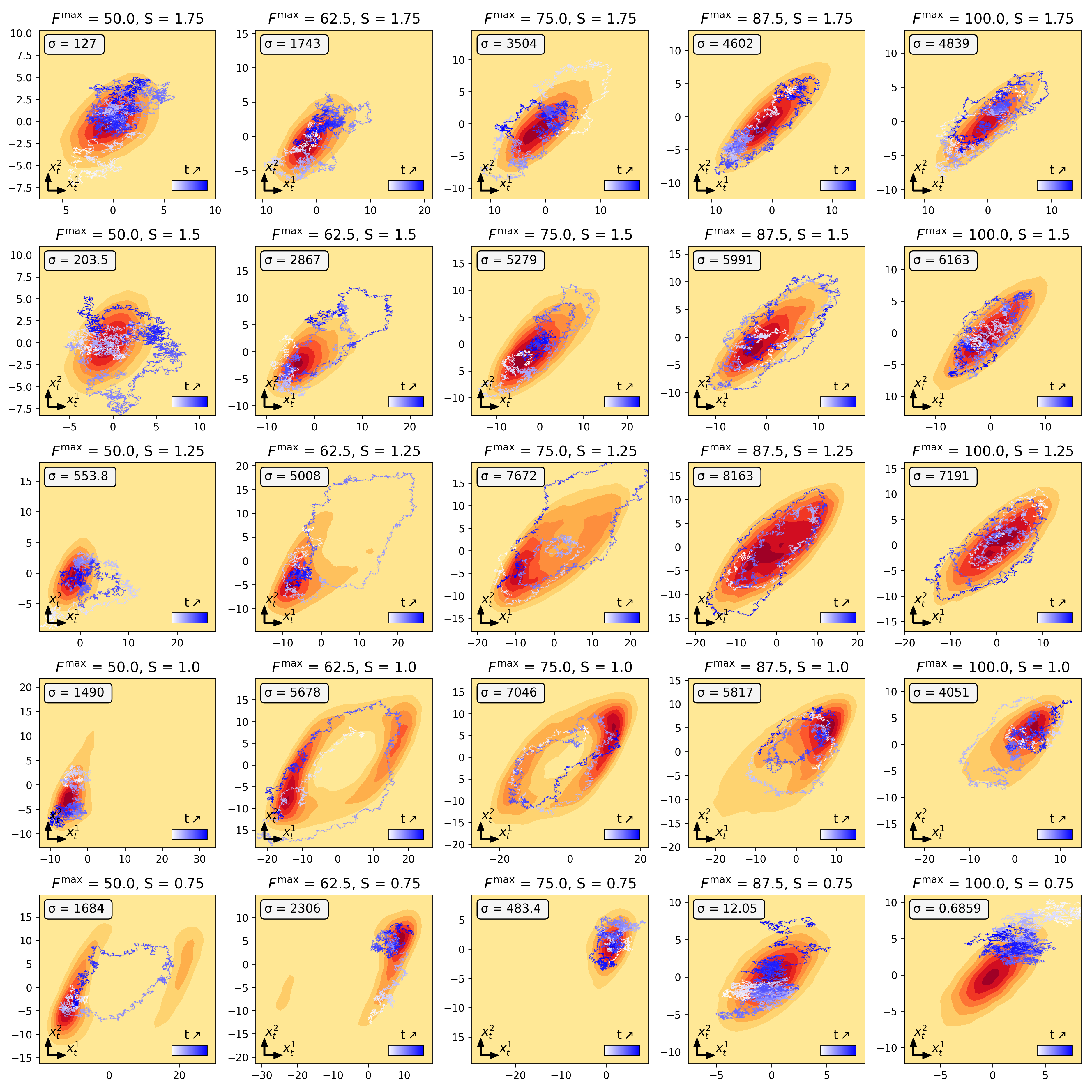}
	\caption{$p(\pmb{x}_t)$ for all analyzed traces generated by the hair cell model, visualized as a contour map. Regions with deeper red shading correspond to higher values of $p(\pmb{x}_t)$. An overlaid segment of a stochastic trace illustrates its path through the probability landscape.}
	\label{FIG_all_pdf_Bull}
\end{figure*}
To calculate the inflow rate $\inflow$, we first estimate the covariance matrix $C_{\nabla \phi}(0)$ of the effective forces $-\nabla \phi(\pmb{x})$ associated with the system. This is done by estimating the joint probability distribution $p(\pmb{x})$ of the observed stochastic trace using non-parametric kernel density estimation. Figures \ref{FIG_all_pdf_lin} and \ref{FIG_all_pdf_Bull} show the estimated probability density functions $p(\pmb{x})$ alongside a segment of the trajectories for all traces analyzed in the paper. Note that, especially for the bullfrog hair cell model, the traces exhibiting a higher degree of circulation correspond to those with a higher entropy production rate $\sigma$. To enhance robustness, the trajectory data is divided into smaller independent subtraces, which are processed individually. For each subtrace, the numerical gradient of the logarithm of the estimated $p(\pmb{x})$ is computed to derive the effective forces. The force covariance matrix $C_{\nabla \phi}(0)$ is computed by averaging the estimated vector field, as shown, for example, in Figures \ref{FIG1}c and \ref{FIG2}c, weighted by the estimated probability density function $p(\pmb{x})$. To enhance the accuracy of the final estimates, the results from all subtraces are averaged. The uncertainty in these estimates is quantified using the standard error.

The inflow rate $\inflow = {\rm Tr}\!\left[\pmb{D}\pmb{C}_{\nabla\phi}(0)\right]$ is then calculated in a straightforward way, and its statistical error is determined using the error propagation formula. Finally, this result is combined with the outcomes of the previous analysis to compute $\sigma$ and its associated statistical error, again using the error propagation formula.

\section{Benchmarking under Measurement Noise}

\tcr{Experimental trajectories are inevitably affected by measurement noise.  Typical sources include Gaussian localization errors, quantization noise from digitization (a uniform rounding error with variance $\Delta^2/12$, where $\Delta$ is the discretization step) and finite exposure times that effectively blur the signal, introducing correlated noise. All these factors introduce additional variability at high frequencies, and the key practical question is how they influence the short-time correlation functions on which our estimator relies.
}

\tcr{A simple and versatile way to represent such disturbances is as an additive process,
\begin{equation}
x_{\mathrm{obs}}(t) = x(t) + \eta(t),
\end{equation}
with 
\begin{align}
    \langle \eta(t) \rangle = 0\,, && D_{\eta} \equiv \langle (\Delta \eta )^2 \rangle = \alpha^2 \,\langle (\Delta x )^2 \rangle \,,
\end{align}
where the parameter $\alpha$ controls the noise level relative to the clean signal. The temporal structure of $\eta(t)$ is captured by its autocorrelation. A particularly convenient model is an exponentially correlated noise,
\begin{equation}
\langle \eta(s)\eta(s+t)\rangle = D_{\eta} e^{-|t|/\tau},
\end{equation}
with correlation time $\tau$. In the limit $\tau \to 0$ this expression reduces to $D_{\eta} \delta_{t,0}$, corresponding to white noise that is uncorrelated from one sample to the next. This formulation therefore covers both short-correlated and effectively white noise sources.}

\tcr{The impact on correlation functions is clear in the white-noise limit. Only the value at $t=0$ is shifted,
\begin{equation}
C_{\mathrm{obs}}(t) =
\begin{cases}
C_x(0)+D_{\eta}, & t=0,\\[4pt]
C_x(t), & t\neq 0,
\end{cases}
\end{equation}
but this shift is enough to strongly bias naïve finite-difference estimates. Indeed, for the first derivative it holds
\begin{equation}
\dot C_{\mathrm{obs}}(0)\;\approx\;\frac{C_{\mathrm{obs}}(dt)-C_{\mathrm{obs}}(0)}{dt}
\;=\;\dot C_x(0) - \frac{D_{\eta}}{dt}\,,
\end{equation}
while the second derivative yields
\begin{equation}
\ddot C_{\mathrm{obs}}(0)\;\approx\;\frac{C_{\mathrm{obs}}(dt)-2C_{\mathrm{obs}}(0)+C_{\mathrm{obs}}(-dt)}{dt^2}
\;=\;\ddot C_x(0) - \frac{2D_{\eta}}{dt^2}\,.
\end{equation}
Both errors diverge as $dt\to 0$, showing why direct differencing at $t=0$ is problematic in the presence of noise.}

\tcr{Our approach is based on fitting correlation functions, which makes it possible to avoid $t=0$ (or short times in general) in presence of noise. Indeed, one can always fit the measured correlation for times $t \geq t_{\min}$ with a small sum of exponentials,
\begin{equation}
C(t)\,\simeq\,\sum_{i=1}^n A_i e^{-t/\tau_i},
\end{equation}
and then obtain the derivatives analytically with a Taylor expansion,
\begin{equation}
\dot C(0) = -\sum_i \frac{A_i}{\tau_i}, 
\qquad 
\ddot C(0) = \sum_i \frac{A_i}{\tau_i^2}.
\end{equation}
The exponential form is flexible enough to reproduce the smooth short-time shape of the true system, while excluding the noise-dominated region ensures that high-frequency perturbations do not contaminate the estimates. The essential requirement is that the noise correlation time $\tau$ is shorter than the intrinsic timescales of the system, so that there is always a window of times where the signal dominates and can be fitted reliably.}

\begin{figure*}[t!]
 	\centering
	\includegraphics[width=1\textwidth]{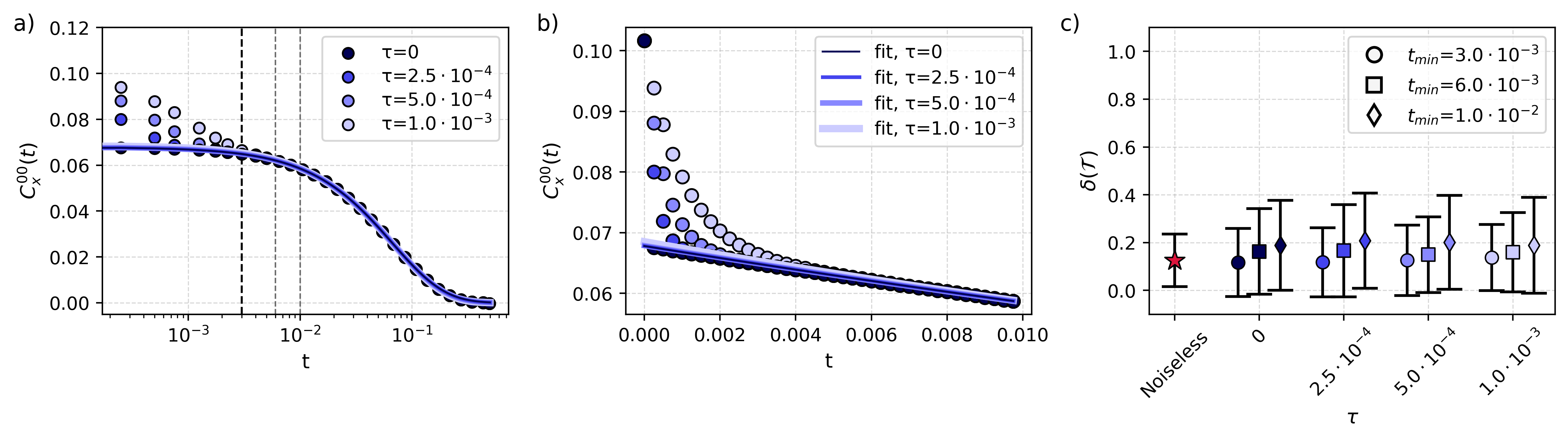}
	\caption{\tcr{(a,b) Autocorrelation functions from noisy trajectories ($\alpha=0.5$) with different noise correlation times $\tau$. Fits with two exponentials (solid lines) accurately reproduce the true short-time structure despite contamination at small $t$. (a) shows a log-scale overview, with vertical dashed lines indicating the chosen values of $t_{min}$; (b) zooms in on early times. (c) Relative error in traffic estimates as a function of noise correlation time and $t_{\min}$, the lower bound of the fit interval. Each point averages 25 trajectories; error bars are standard deviations. The red star is the baseline estimate from noiseless data fitted over $[0,t_{\max}]$. Performance degrades slightly with increasing $\tau$ or larger $t_{\min}$, but remains accurate and free of the $1/dt^2$ sensitivity of naïve differencing.}}
	\label{linear_fit_bench}
\end{figure*}

\tcr{We first illustrate this with the Brownian gyrator. For each coordinate, we corrupt the original trajectories as
\begin{equation}
x^{i}_{\mathrm{obs}}(t)=x^{i}(t)+\eta^{i}(t),
\qquad \langle \eta^i(t) \rangle = 0,
\end{equation}
where the added noise has correlation
\begin{equation}
\langle \eta^i(t)\,\eta^j(t') \rangle
= D_{\eta}^i \delta_{ij}\, e^{-|t-t'|/\tau},
\end{equation}
where $D^i_{\eta} = \alpha^2 \, \langle (\Delta x^i)^2\rangle $ so that $\alpha$ sets the level of the noise, and $\tau$ is the noise correlation time.  
We then vary both the correlation time $\tau$ and the starting point $t_{\min}$ of the fit to test the robustness of the fits in various situations. Each correlation function is fitted with two exponentials, 
\begin{equation}
C(t) \simeq A_1 e^{-t/\tau_1} + A_2 e^{-t/\tau_2},
\end{equation}
and the derivatives $\dot C(0)$ and $\ddot C(0)$ are obtained analytically from the fitted parameters. 
Because the gyrator is analytically solvable, the estimated traffic $\overline{\mathcal{T}}$ can be directly benchmarked against the exact value $\mathcal{T}$. 
As shown in Fig.~\ref{linear_fit_bench}a)-b), the correlated noise distorts the measured $C_{\mathrm{obs}}(t)$ near the origin, but the fitted form extrapolates correctly to $t=0$ even when the fit starts at finite $t_{\min}$. 
Panel (c) quantifies this effect: the relative error 
\begin{equation}
\delta(\overline{\mathcal{T}}) = \frac{|\overline{\mathcal{T}} - \mathcal{T}|}{|\mathcal{T}|}
\end{equation}
is averaged across 25 independent trajectories for different combinations of model parameters (as in the main text), and the error bars represent the corresponding standard deviations. 
The red star in Fig.~\ref{linear_fit_bench}c) is the baseline estimate from noiseless data fitted over $[0, t_{\max}]$. 
All other estimates remain close to this reference, showing only modest degradation as either $\tau$ or $t_{\min}$ increase. 
Importantly, there is no divergence with the sampling interval $\mathrm{d}t$, confirming that the fitting procedure suppresses the $1/(\mathrm{d}t)^2$ sensitivity of naïve differencing and is robust to correlated measurement noise. 
The near constancy of the relative error across $\tau$ also indicates that the method remains stable as long as the noise correlation time is shorter than the intrinsic relaxation times of the system (in our case $\sim 10^-2$ for the chosen parameters). Finally, a lower sampling rate, corresponding to a larger $\mathrm{d}t$, has a similar effect to increasing $t_{\min}$: it reduces the number of short-time points but does not bias the fitted derivatives, leading only to slightly larger uncertainties due to fewer data constraints at early times.}

\begin{figure*}[t!]
 	\centering
	\includegraphics[width=0.95\textwidth]{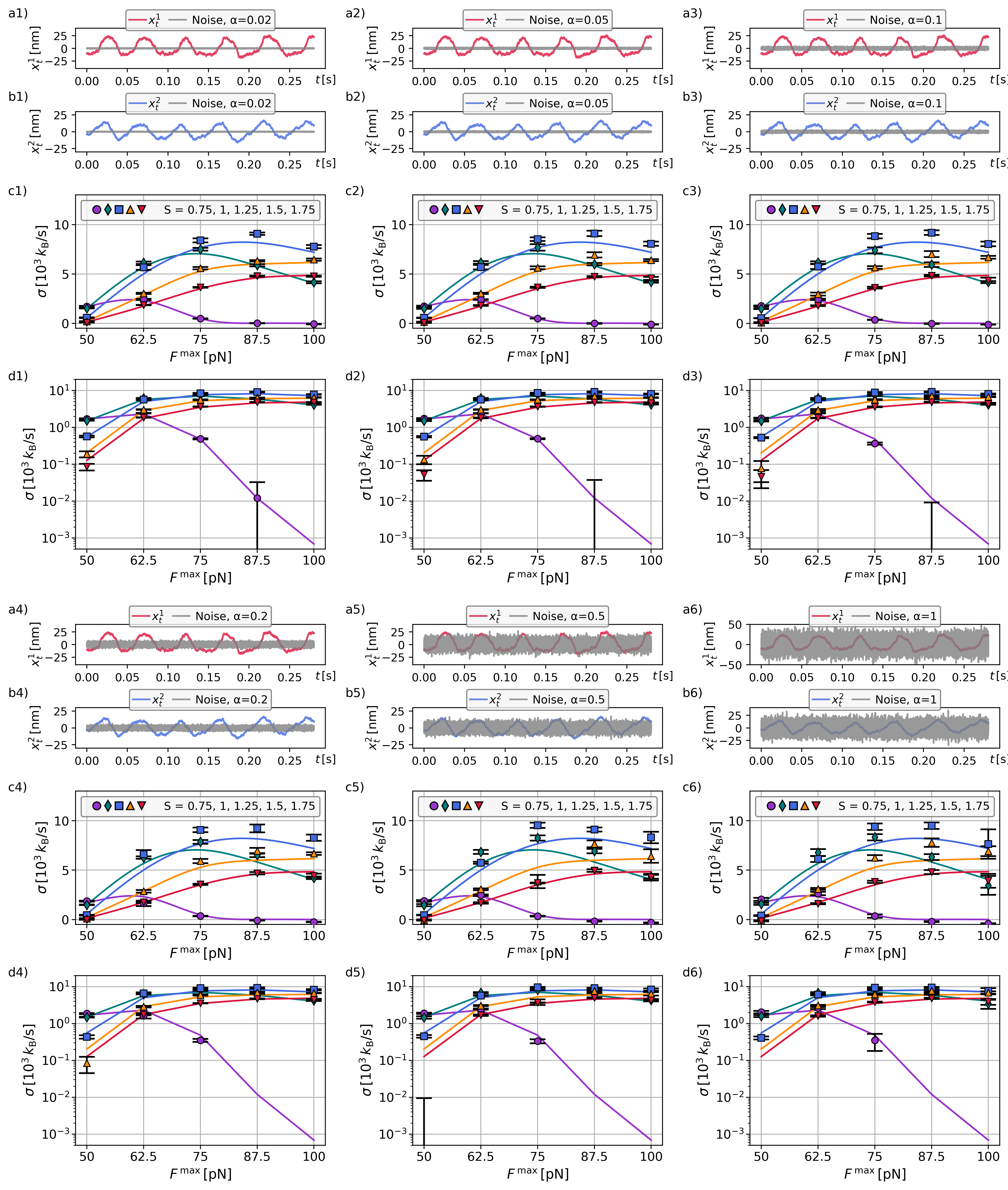}
	\caption{\tcr{Benchmark of entropy production estimation for the hair-bundle model under white noise.
    Panels (a,b) show representative trajectories with added noise at different levels $\alpha$. (c) Estimated entropy production $\sigma$ as a function of the control parameter $F^{\max}$ for several values of $S$ (colors). Solid lines are ground-truth $\sigma$, markers are estimates with uncertainties. Rows correspond to increasing $\alpha$, up to $\alpha=1$. Even at this level, estimates remain reliable, especially at high dissipation.}}
	\label{noise_bench}
\end{figure*}

\tcr{We next examine the nonlinear hair-bundle model, where no analytical estimates of the traffic are available, and therefore the benchmark is performed directly on the entropy production rate $\sigma$ obtained from Eq. \eqref{sekimoto}. 
The trajectories are perturbed in an analogous way,
\begin{equation}
x^{i}_{\mathrm{obs}}(t)=x^{i}(t)+\eta^{i}(t),
\qquad \langle \eta^i(t) \rangle = 0,
\end{equation}
but now with white measurement noise
\begin{equation}
\langle \eta^i(t)\,\eta^j(t') \rangle
= D_{\eta}^i\, \delta_{ij}\, \delta_{t,t^{\prime}}\,,
\end{equation}
where again $D^i_{\eta} = \alpha^2 \, \langle (\Delta x^i)^2\rangle$. Here we vary $\alpha$ up to $\alpha=1$.
Since white noise only shifts the zero-time correlation, we exclude the point at $t=0$ from the fit and start at $C(\mathrm{d}t)$.  
The added noise primarily affects the inflow contribution, biasing the estimates to smaller values because larger variances correspond to weaker effective forces. This can be seen in Fig.~\ref{noise_bench} where estimates of traces with a low $\sigma$, become slightly negative (hence not appearing in log-scaled plots). However, for $\sigma \gtrsim 100\,k_{\mathrm{B}}/{\rm s}$ the method remains very accurate even at the highest noise levels ($\alpha=1$). 
These results confirm that, although measurement noise can hinder the detection of small dissipation rates, the estimator robustly recovers $\sigma$ from very noisy measurements of strongly driven regimes.}

\tcr{In summary, these tests demonstrate that the estimator is robust to measurement perturbations. The flexibility of exponential fitting ensures that short-correlated or white noise is suppressed by excluding $t=0$, while exponentially correlated noise is accommodated by starting the fit at a slightly larger $t_{\min}$. As long as the noise timescale is shorter than the characteristic relaxation times of the system, the true short-time dynamics are recovered and the kinetic quantities of interest, especially the traffic, can be inferred without bias. In realistic experimental conditions, one also has some a priori knowledge about the setup (for instance, whether localization noise or exposure blur is present) and this knowledge should be used to choose the most appropriate fitting window. Exploiting such prior information makes the method not only robust, but also adaptable to the specifics of each experimental situation.}

\end{document}